\documentclass[a4paper,fleqn]{cas-dc}

\usepackage[numbers,sort&compress]{natbib}
\usepackage{subcaption}
\usepackage{amsmath}
\usepackage{amssymb}
\usepackage{arydshln}
\usepackage{mathtools}
\usepackage{epstopdf}
\usepackage{soul}
\usepackage{afterpage}
\def\tsc#1{\csdef{#1}{\textsc{\lowercase{#1}}\xspace}}
\tsc{WGM}
\tsc{QE}
\begin{document}
\let\WriteBookmarks\relax
\def\floatpagepagefraction{1}
\def\textpagefraction{.001}

\shorttitle{\textit{GaAs-Based Near-Field Thermophotonic Devices: Approaching The Idealized Case With One-Dimensional PN Junctions}}    

\shortauthors{Legendre, Chapuis}  

\title [mode = title]{GaAs-Based Near-Field Thermophotonic Devices: Approaching The Idealized Case With One-Dimensional PN Junctions}  

\author[1]{Julien Legendre}[orcid=0000-0001-7316-1954]

\cormark[1]

\ead{julien.legendre@insa-lyon.fr}

%\ead[url]{<URL>}

%\credit{<Credit authorship details>}

\author[1]{Pierre-Olivier Chapuis}


\affiliation[1]{organization={Univ Lyon, CNRS, INSA-Lyon, Université Claude Bernard Lyon 1, CETHIL UMR5008},
            %addressline={9 rue de la Physique}, 
            postcode={F-69621}, 
            city={Villeurbanne},
%          citysep={}, % Uncomment if no comma needed between city and postcode
            %state={},
            country={France}}

\cortext[1]{Corresponding author}

\begin{abstract}
Thermophotonics (TPX) is a technology close to thermophotovoltaics (TPV), where a heated light-emitting diode (LED) is used as the active thermal emitter of the system. It allows to tune the heat flux, by means of electroluminescence, to a spectral range matching better the gap of a photovoltaic cell. The concept is extended to near-field thermophotonics (NF-TPX), where enhanced energy conversion is due to both electric control and wave tunneling. We perform a thorough numerical analysis of a GaAs-based NF-TPX device, by coupling a near-field radiative heat transfer solver based on fluctuational electrodynamics with an algorithm based on a simplified version of the drift-diffusion equations in 1D. This allows for the investigation of the emission and absorption profiles in the LED and the photovoltaic (PV) cell, and for the scrutiny of the impact of key parameters. We also demonstrate that the performance obtained with this algorithm can approach idealized cases for improved devices. For the considered simplified architecture and 300 K temperature difference, we find a power density output of 1 W.$\text{cm}^{-2}$, underlining the potential for waste heat harvesting close to ambient temperature.
\end{abstract}

% Keywords
\begin{keywords}
Near-field radiative heat transfer \sep 
Thermophotonics \sep
Light-emitting diodes \sep
Photovoltaics \sep
GaAs \sep
\end{keywords}

\maketitle

% Main text
\section{Introduction}\label{Intro}
The necessity of a change in our relation with energy needs no longer to be proven. This change has to happen on several important topics: the generation of clean energy, the reduction of consumption of machines and buildings, and the recuperation of energy losses. On this last subject, recent solid-state heat engines avoid notably (potentially polluting) refrigerants and moving parts (thus vibrations). Thanks to these benefits, solid-state heat engines can be interesting candidates for both terrestrial and spatial applications.\newline
Several of these engines have seen good recent development in the last 20 years thanks to dynamic research \cite{Green2016,Melnick2019}: we can cite thermoelectric devices \cite{Chen2018a,Mao2018}, thermophotovoltaic (TPV) devices \cite{Daneshvar2015}, thermionic devices \cite{Schwede2010,Campbell2021} or hot-carrier cells \cite{Konig2010,Li2019}. In the TPV field, if a large fraction of the research focuses on the development of improved emitters \cite{Sakakibara2019} or cell rear-face reflectors \cite{Fan2020}, advanced concepts are also under development \cite{Tervo2018}, including hybridization with thermionic devices \cite{Datas2016,Zeneli2020}, near-field (NF) enhancement \cite{Whale2002,Park2008,Francoeur2011} and thermophotonics (TPX) \cite{Harder2003,Zhao2019}. In this work, we aim to study the combination of these last two concepts through the numerical analysis of a near-field thermophotonic (NF-TPX) device \cite{Zhao2018}.\newline
In a near-field thermophotovoltaic (NF-TPV) device, by decreasing the gap distance between the emitter and the photovoltaic (PV) cell to the order of hundreds of nanometers or below, evanescent waves start to participate to the heat transfer along with propagative modes \cite{Zhang2007}. Heat transferred to the PV cell can then be increased by a factor 10 to even 100 for small gap distances \cite{Francoeur2011}, exceeding therefore the blackbody limit and increasing by a similar order of magnitude the electrical power produced by the PV cell. However, the emitter cannot be controlled dynamically. The practical capabilities of such device have already been demonstrated \cite{Fiorino2018,Inoue2019,Bhatt2020,Lucchesi2021,Mittapally2021,Inoue2021a}.\newline
% Figure 1: system
\begin{figure}[]
	\centering
	\includegraphics[scale=0.4]{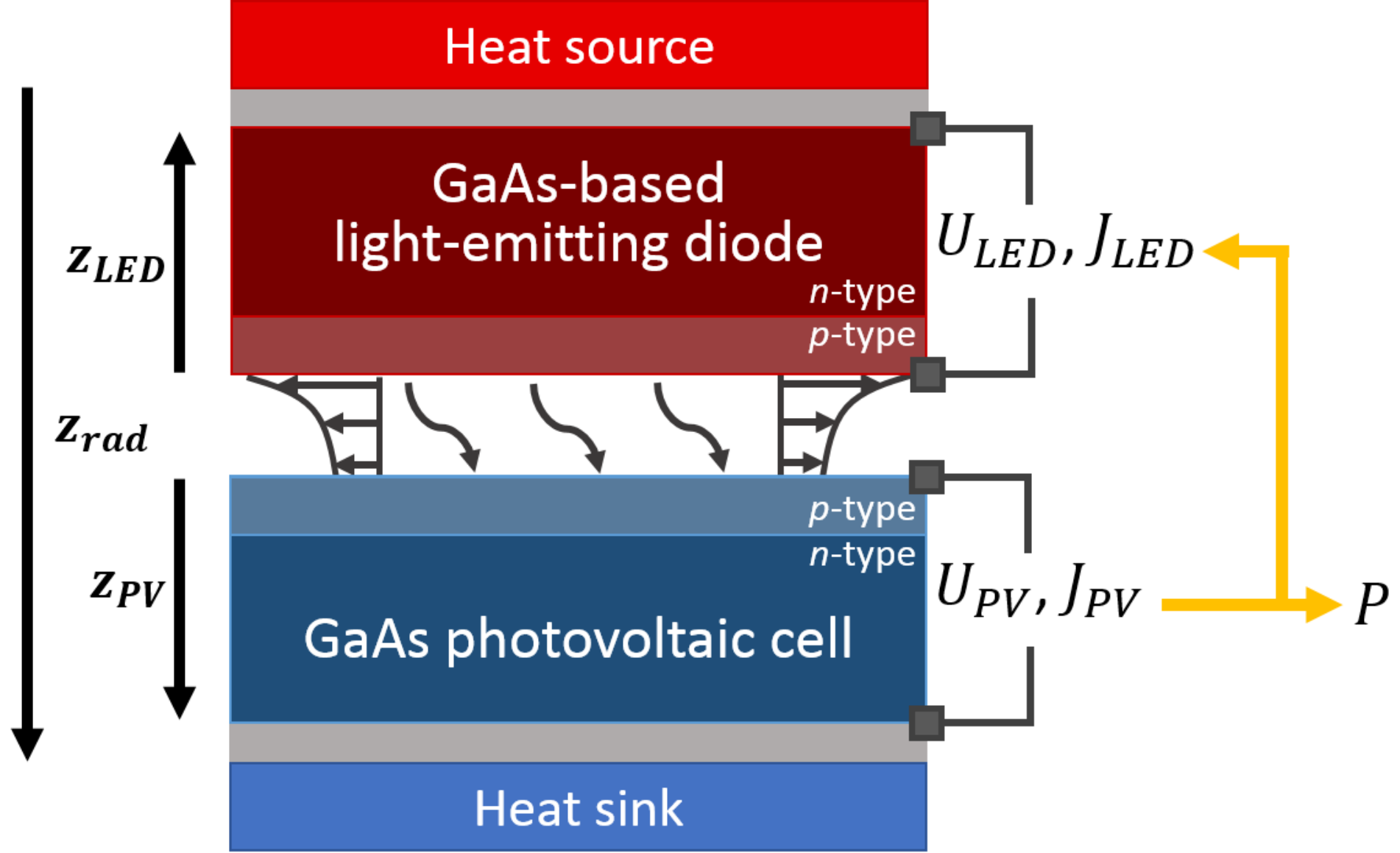}
	  \caption{Schematic of a GaAs near-field thermophotonic device. The LED is maintained at 600 K, while the PV cell is kept at 300 K. In the nanogap, propagative and evanescent modes participate to the radiative heat transfer. Absorption of the radiation by the photovoltaic cell allows for the production of electrical power, a fraction of which is redirected towards the light-emitting diode for enhancement of the radiative heat transfer by electroluminescence. Ideal mirrors (in grey) allow for the confinement of photons inside the device.}\label{fig1}
\end{figure}Thermophotonic devices rely on the recent development of light-emitting diodes (LED). Thanks to the electroluminescent effect, the emission spectrum of the LED can be tuned by the application of a voltage, and can also exceed the blackbody limit. If most of this development was directed towards lighting applications, recent research has pointed out the interest of LEDs for thermodynamic applications as an electroluminescent refrigerator/heat pump \cite{Oksanen2010,Sadi2019a,Sadi2020} or as a key part of a TPX device. In the latter, the LED is placed in front of a PV cell and is heated above ambient temperature (in a heat engine configuration). Thanks to the electroluminescence effect, the radiative power transferred to the PV cell can be increased by order of magnitudes compared to the TPV case \cite{Harder2003}. However, the LED is consuming electrical power: a TPX device can therefore produce electrical power only when made of high-efficiency LED and PV cell, this last argument being the reason of the recent interest in the device. Indeed, with the maturation of the LED technology, the achievable efficiency has increased largely in the last decades. Among the well-known semiconductors regularly used for PV and TPV applications, gallium arsenide (GaAs) junctions has reached extremely high internal and external quantum efficiencies (above 99\% \cite{Bender2013a}), making it an ideal candidate for TPX application as underlined by the Aalto group \cite{Sadi2020}, even with a larger bandgap energy than what is usually used for TPV (i.e. closer to visible than to mid-infrared).\newline
By combining near-field effects with electroluminescent enhancement in a NF-TPX device, the radiative heat transfer increase is coupled and permits a dramatic increase of the electrical power output. Due to the limited temperatures current LEDs can withstand (600 to 700 K, see the review on high-temperature pn junctions in \cite{Vaillon2020}), this kind of device is ideal for low-grade heat recovery application. This could allow for a photonic device to be performant in a range where standard TPV is barely used \cite{Daneshvar2015,Sakakibara2019,Fan2020} because of its poor efficiency \cite{Green2016}, being then competitive against thermoelectrics, a technology commonly used in this range of temperature \cite{Mao2018}.\newline
In the first article on NF-TPX written by Zhao \textit{et al.} \cite{Zhao2018}, a 1D near-field radiative heat transfer (NF-RHT) model was coupled with a 0D detailed balance (DB) model for the computation of the currents. We propose here to analyze the performance of a GaAs NF-TPX device, by coupling a similar NF-RHT code with an improved charge carrier behaviour model relying on the solution of the simplified drift-diffusion (SDD) equations in 1D, often used in standard PV studies \cite{Nelson2003,Kitai2011}.

\section{Modelling}\label{Mod}
The NF-TPX device considered is shown in Figure \ref{fig1}. The PV cell is a GaAs pn junction, while the LED is made of a GaAs-based alloy (this matter is discussed at the end of Section \ref{Mod} introduction). The p- and n-regions are respectively the front and back layers of the two junctions. The LED is heated at 600 K, which allows for large radiative heat transfer while keeping the quantum efficiency high \cite{Maros2015,Sun2017}. The PV cell is kept at ambient temperature (300 K). On the back surface of the device, two perfect mirrors (i.e., with a dielectric function with an infinitely large real part and a null imaginary part) allows for the confinement of the photons inside the device. The main device parameters are shown in Table \ref{ParamDevice}. On the right of Figure 1 are provided the main electrical parameters, where $U$ is the voltage and $J$ the current density. $P$ represents the electrical power density produced by the device.\newline
% Table : Device parameters
\begin{table}[<options>]
\caption{Device parameters. The first value given is related to the LED, the second one to the PV cell. Opt. 1 refers to the optimized values obtained with surface recombinations, Opt. 2 to those obtained without surface recombinations (see Section \ref{ImprovingGeo}).}\label{ParamDevice}
\begin{tabular*}{\tblwidth}{@{}LLLLLL@{}}
\toprule
Parameter & Unit & Symbol & Ref. & Opt. 1 & Opt. 2 \\ % Table header row
\midrule
Temperature & K & $T$ & 600 &&\\
                  &&& 300 &&\\
\hdashline
P-region doping & $\text{cm}^{-3}$ & $N_{a}$ & $10^{20}$ &&\\
                                           &&& $10^{20}$ &&\\
\hdashline
N-region doping & $\text{cm}^{-3}$ & $N_{d}$ & $10^{18}$ &&\\
                                           &&& $10^{18}$ &&\\ 
\hdashline
Gap distance & nm & \textit{d} & 10 &&\\
\hdashline
Mirror thickness & nm & $t_m$ & 10 &&\\
\hdashline
P-region thickness & nm & $t_{p}$ & 400 & 10 & 10\\
                                &&& 400 & 10 & 10\\
\hdashline
N-region thickness & nm & $t_{n}$ & $10^{4}$ & $10^{4}$ & 500 \\
                                &&& $10^{4}$ & $10^{4}$ & $10^{3}$ \\
\bottomrule
\end{tabular*}
\end{table}The modelling process can be divided into two main steps:
\begin{enumerate}[a)]
\item calculation of the near-field radiative heat transfer (including both below and above-bandgap photons, see Section \ref{NFRHT})
\item calculation of the I-V characteristic through the modelling of the charge carrier behaviour (see Section \ref{CarrierBehaviour})
\end{enumerate}
The different modelling parameters used in this work can be found in Table \ref{ParamMod}. By taking into account the variation of GaAs bandgap with temperature \cite{Gonzalez-Cuevas2007}, the bandgap of a GaAs LED goes down to 1.28 eV, which has a significant impact on performances due to bandgap mismatch with the PV cell ($E_g^{\text{300 K}}=1.42$ eV). Instead of pure GaAs, the LED is therefore made of a GaAs-based alloy, which displays a bandgap of 1.42 eV at 600 K. In addition, optical properties of such alloys are very close to those of GaAs at 300 K (precisely because the bandgaps are matched). This allows us to model the performance of a GaAs-based LED using well-known GaAs optical properties at 300 K and electrical properties at 600 K. In this paper, we indistinctly refer to the device as 'GaAs NF-TPX' or 'GaAs-based NF-TPX', the LED being always made of GaAs-based alloys to keep its bandgap matched with that of the PV cell.

% Table : Modelling parameters
\begin{table}[<options>]
\caption{Modelling parameters.}\label{ParamMod}
\begin{tabular*}{\tblwidth}{@{}LLLLR@{}}
\toprule
Parameter & Unit & Symbol & Value & Reference  \\ % Table header row
\midrule
Bandgap energy & eV & $E_{g}$ & 1.42 & \cite{Adachi1988}\\
Dielectric function & - & $\varepsilon$ & & \\
\hphantom{abc}\textit{Interband} & & & & \cite{Gonzalez-Cuevas2007} \\
\hphantom{abc}\textit{Drude}\textit{-Lorentz} & & & & \cite{Adachi1994,Losego2009} \\
Static rel. permittivity & - & $\varepsilon_{s}$ & 12.9 & \cite{IoffePhysico-TechnicalInstitute} \\
Electron rel. eff. mass & - & $m_{e}^{*}$ & 0.063 & \cite{IoffePhysico-TechnicalInstitute}\\
Hole rel. eff. mass & - & $m_{h}^{*}$ & 0.53 & \cite{IoffePhysico-TechnicalInstitute} \\
Electron mobility & $\text{cm}^{2}.\text{V}^{-1}.\text{s}^{-1}$ & $\mu_{e}$ & & \cite{Sotoodeh2000} \\
Hole mobility & $\text{cm}^{2}.\text{V}^{-1}.\text{s}^{-1}$ & $\mu_{h}$ & & \cite{Sotoodeh2000} \\
SRH recomb. coeff. & $\text{s}^{-1}$ & $A$ & 3$\times 10^{5}$ & \cite{Sadi2019a} \\
%Radiative recombination coeff. B ($\text{cm}^{3}\text{s}^{-1}$) & $2 \times 10^{-10}$ & \cite{Sadi2019a} \\
Auger recomb. coeff. & $\text{cm}^{6}.\text{s}^{-1}$ & $C$ & $10^{-30}$ & \cite{Sadi2019a} \\
Surf. recomb. coeff. & $\text{m}.\text{s}^{-1}$ & $S_{n}$ & 5$\times 10^{2}$ & \cite{Blandre2017} \\
\bottomrule
\end{tabular*}
\end{table}

\subsection{Near-field radiative heat transfer}\label{NFRHT}
The near-field radiative heat transfer between the LED and the PV cell is simulated using the Fluctuational Electrodynamics theory developed by Rytov \cite{Rytov1989}, which allows for an accurate description of both propagative, frustrated and surface modes, the two latter being evanescent modes that are not described by the macroscopic framework. The random movement of charges at any point, characterized by the Fluctuation-Dissipation Theorem, can be linked to the electromagnetic wave perceived at any other point through Green tensors. Then, the radiated power is simply obtained as the mean value of the Poynting vector. By using Francoeur’s formalism \cite{Francoeur2009}, the spectral photon flux density $\gamma_{\omega}$, which is related to the spectral heat flux density $q_{\omega}$ as $q_{\omega}=\hbar\omega \gamma_{\omega}$, emitted at a layer \textit{s} and received at a position $z_{l}$ in layer \textit{l} can be expressed as
\begin{equation}\label{PhotonFlux}
    \gamma_{\omega,sl}(z_{l})=\left( n^{0}_{s}-n^{0}_{l} \right) \mathcal{T}_{sl}(z_{l})=\gamma_{\omega,sl}^{in}(z_{l})-\gamma_{\omega,sl}^{out}(z_{l}).
\end{equation}

In this expression, $n^{0}$ is the modified Bose-Einstein distribution, which is tuned by means of electroluminescence through the value of the electrochemical potential $\mu$ (defined as the difference between the electron and hole quasi Fermi levels $\mu=E_{Fn}-E_{Fp}$):
\begin{equation}\label{ModifiedBoseEinstein}
    n^{0}(\omega,\mu,T)=\begin{cases}
        \text{exp}\left(\hbar\omega/\left(k_{B}T\right)-1\right)^{-1}, &\hbar\omega < E_{g} \\
        \text{exp}\left(\left(\hbar\omega-\mu\right)/\left(k_{B}T\right)-1\right)^{-1}. &\hbar\omega \geq E_{g} \\
    \end{cases}
\end{equation}

$\textit{E}_{g}$ is the gap energy, $\hbar$ is the reduced Planck constant, $\omega$ the radiation angular frequency, $k_{B}$ the Boltzmann constant and \textit{T} the temperature.\newline
In this work, we consider that the electrochemical potential $\mu$ is constant and equal to $eU$, a common approximation whose precision has been studied in \cite{Callahan2021}. By doing this approximation, we assume that the Bose-Einstein distribution is similar in all points of the LED (resp. PV cell), therefore that the net photon flux density between two points of the same device is null.
\newline
The other factor, $\mathcal{T}_{sl}(z_{l})$, corresponds to an equivalent transmission coefficient between the layer \textit{s} and any point of the layer \textit{l}, and can be obtained from the transmission coefficient $\mathcal{T}_{sl}(z_{s},z_{l})$ between two points positioned respectively in layers \textit{s} and \textit{l}
%\begin{figure*}
%\hrule
\begin{subequations}\label{TransmissionCoeff}
    \begin{gather}
        \mathcal{T}_{sl}(z_{l})=\int_{z_{s}} \mathcal{T}_{sl}(z_{s}^{'},z_{l}) \,dz_{s}^{'}\\
        \mathcal{T}_{sl}(z_{s},z_{l})=\frac{k_{v}}{\pi^{2}}\Re{\left( i\varepsilon_{r1}^{''}\int_{0}^{+\infty} F_{\omega,sl}(k_{\rho}) k_{\rho} \,dk_{\rho}\right) }\\
        F_{\omega,sl}(k_{\rho})=\begin{bmatrix} \phantom{+}g_{sl\rho\rho}^{E}(\omega,k_{\rho},z_{s},z_{l})g_{sl\theta\rho}^{H*}(\omega,k_{\rho},z_{s},z_{l}) \\+g_{sl\rho z}^{E}(\omega,k_{\rho},z_{s},z_{l})g_{sl\theta z}^{H*}(\omega,k_{\rho},z_{s},z_{l})
        \\-g_{sl\theta\theta}^{E}(\omega,k_{\rho},z_{s},z_{l})g_{sl\rho\theta}^{H*}(\omega,k_{\rho},z_{s},z_{l}) \end{bmatrix}.
    \end{gather}
\end{subequations}
%\end{figure*}

The terms \textit{g} correspond to the different Weyl components of the Green tensors, $k_{\rho}$ is the parallel component of the wavevector $\textbf{k}$, $k_{v}$ is the vacuum wavenumber corresponding to the given frequency $\omega$ and $\varepsilon ''$ is the imaginary part of the dielectric function $\varepsilon$.\newline
In this work, we assume that the LED and PV cell temperatures are kept constant; thus, below-bandgap photons do not have any impact on the electrical performance, only on the device efficiency. In the following, the photon flux density $\gamma_{sl}$ used in the electrical models is therefore defined as the integral of its spectral quantity over above-bandgap photons: $\gamma_{sl}=\gamma_{\hbar\omega\geq E_{g},sl}=\int_{E_{g}/\hbar}^{+\infty} \gamma_{\omega,sl} \,d\omega$.

\subsection{Charge carrier behaviour}\label{CarrierBehaviour}
Once the spectral photon flux density $\gamma_{\omega}$ is obtained, the charge carrier behaviour can be modeled. In order to do so, we use two different methods:
\begin{enumerate}[a)]
\item The detailed balance (DB) method (see Section \ref{DetBal}) allows to easily obtain the current thanks to a balance between photons and charge carriers. This method has been used in the literature for the computation of the currents in a TPX device \cite{Harder2003,Zhao2018,Sadi2020}.
\item The simplified drift-diffusion (SDD) method, in which the drift-diffusion equations are solved using standard approximations, is routinely used in photovoltaics \cite{Nelson2003,Kitai2011} but has not been used for TPX devices until now.
\end{enumerate}
The use of these two methods allows for a complete description of the device performance. By using the DB method, we can estimate the ideal capabilities of such a system; then, thanks to the SDD method, the power output of a specific device can be obtained and its efficiency estimated in comparison to the ideal case.

\subsubsection{Detailed balance (0D)}\label{DetBal}
The photon flux density is known at any point of the system. In order to obtain a direct relation with the current densities, several assumptions have to be made \cite{Harder2003}:
\begin{enumerate}[a)]
\item \textbf{absorption} - each above-bandgap photon absorbed generates a unique electron-hole (e-h) pair: $\Tilde{G}_{rad}=\gamma_{abs}=\gamma_{sl}^{in}(z_l^-)-\gamma_{sl}^{in}(z_l^+)$ (defining $\Tilde{x}=\int x \,dz$).
\item \textbf{radiative recombination} - photon recycling is neglected, thus the local radiative recombination corresponds to the photon flux density emitted towards the other device: $\Tilde{R}_{rad}=\gamma_{em}=\gamma_{sl}^{out}(z_l^-)-\gamma_{sl}^{out}(z_l^+)$.
\item \textbf{non-radiative recombination} - the level of non-radiative recombinations and of thermal generation of e-h pairs is set by the Internal Quantum Efficiency (IQE): $\Tilde{R}_{nonrad}=(1/\text{IQE}-1)\gamma_{em}$ and $\Tilde{G}_{nonrad}=(1/\text{IQE}-1)\gamma_{em}(U_l=0)$.
\end{enumerate}
Since $J=e(\Tilde{G}-\Tilde{R})$, we obtain
\begin{equation}\label{CompleteJ0D}
    \begin{split}
        J_{l} & = e \Bigl[ \gamma_{abs}-\gamma_{em}-\left(\frac{1}{\text{IQE}}-1 \right)\left(\gamma_{em}-\gamma_{em}(U=0)\right) \Bigr] \\
        & = \!\begin{multlined}[t] e \Bigl[ \left(\gamma_{sl}(z_{l}^{-})-\gamma_{sl}(z_{l}^{+})\right) -\left(\frac{1}{\text{IQE}}-1\right) \Bigr.\\ \Bigl. \left(n^{0}_{l}-n^{0}_{l}(U_{l}=0)\right)\left( \mathcal{T}_{sl}(z_{l}^{-})-\mathcal{T}_{sl}(z_{l}^{+}) \right) \Bigr].\end{multlined}
    \end{split}
\end{equation}
This expression can be simplified if $\text{IQE}=1$:
\begin{equation}\label{PerfectJ0D}
    J_{l}=e \left(\gamma_{sl}(z_{l}^{-})-\gamma_{sl}(z_{l}^{+})\right).
\end{equation}
These two equations are developed in \cite{Harder2003} and \cite{Sadi2020}. Note that in \cite{Zhao2018}, a more thorough model has been developed. However, in the case of this work, we aim to obtain a simple and ideal (or quasi-ideal if $\text{IQE}<1$) model, therefore refining the DB model is not of interest for us.

Note that when using the DB method, we consider that both devices (labeled respectively \textit{s} and \textit{l}) are homogeneous semi-infinite media for the radiative computation, i.e. that $\mathcal{T}_{sl}(z_{l}^{+})\approx 0$ (thus $\gamma_{sl}(z_{l}^{+})\approx 0$). This allows to obtain ideal or quasi-ideal performances which are independent of the LED and PV cell geometries considered. In addition, when considering quasi-ideal cases, the IQE chosen is identical for the LED and the PV cell.

\subsubsection{Drift-Diffusion (1D)}\label{DDE}

For a thorough modelling of a TPX system, Poisson, continuity and drift-diffusion equations should be solved simultaneously in order to obtain the local charge carriers distribution $n$ and $p$, electric field inside the junction $E$ and electron and hole currents $J_{n}$ and $J_{p}$. In 1D, these equations are expressed as
\begin{equation}\label{Poisson}
    \frac{dE}{dz}(z)=-\frac{e}{\varepsilon_{s}}\left( n(z)-p(z)+N_{a}(z)-N_{d}(z) \right)
\end{equation}
\begin{subequations}\label{Continuity}
    \begin{gather}
        \frac{dJ_{n}}{dz}(z)=e(R(z)-G(z)) \\
        \frac{dJ_{p}}{dz}(z)=-e(R(z)-G(z))
    \end{gather}
\end{subequations}
\begin{subequations}\label{DriftDiffusion}
    \begin{gather}
        J_{n}(z)=e\cdot n(z)\mu_{n}E(z)+eD_{n}\frac{dn}{dz}(z) \\
        J_{p}(z)=e\cdot p(z)\mu_{p}E(z)-eD_{p}\frac{dp}{dz}(z).
    \end{gather}
\end{subequations}

These equations being coupled, solving the complete problem would need to implement an iterative process \cite{Gummel1964,Blandre2017}. However, standard approximations are regularly used to simplify the problem \cite{Nelson2003,Kitai2011}:
\begin{enumerate}[a)]
\item the depletion approximation, in which the electric field is analytically defined and becomes independent of the illumination.
\item the low-injection approximation, in which we assume that the illumination is low enough to have constant majority carrier densities (equal to the doping level).
\end{enumerate}

By using these approximations, the problem can be decoupled, obtaining one differential equation for the minority carrier density to be solved in the quasi-neutral regions on the p-side and on the n-side. This differential equation is expressed for electrons in the p-region as
\begin{equation}\label{StandartApprox}
    D_{n}\frac{d^{2}\Delta n_{p}}{dz^{2}}(z_{p})-\frac{\Delta n_{p}(z_{p})}{\tau_{non-rad,n}}+G(z_{p})-R_{rad}(z_{p})=0.
\end{equation}

In this equation, $\Delta n_{p}$ represents the variation of the electron density in the p-region compared to the equilibrium value $n_{p0}$. $z_{p}$ corresponds to the position in the p-region, and the differential equation should be solved between the limit of the depletion region in the p-region $z_{p,dp}$ and the boundary of the region $t_{p}$. The boundary conditions are
\begin{subequations}\label{StandartApproxLimit}
    \begin{gather}
        \Delta n_{p}(z_{p,dp})=n_{p0} \left( \text{exp}\left( \frac{eU}{k_{B}T} \right) -1 \right) \\
        D_{n}\frac{d\Delta n_{p}}{dz}(t_{p})=-S_{n}\Delta n_{p}(t_{p}).
    \end{gather}
\end{subequations}

Note that the differential equation and the boundary conditions are identical for holes in the n-region. $S_{n}$ corresponds to the surface recombination coefficient for electrons at the front interface (see Table \ref{ParamMod}). We consider equivalent surface recombinations at the front and back interfaces of the junctions: $S_{p}=S_{n}=5\times 10^2$ $\text{m.s}^{-1}$. $D_{n,p}$ is the diffusion coefficient, obtained from the mobility $\mu_{n,p}$ through Einstein's equation:
\begin{equation}\label{DiffCoeff}
    D_{n,p}=\mu_{n,p}\cdot k_{B}T/e.
\end{equation}

For the SDD method, as for the DB method, the generation of e-h pairs is computed from the results of the near-field radiative heat transfer model. Here, the two layers (p- and n-regions) of each device are considered for the radiative and charge transport computations, as shown in Figure \ref{fig1}. In order to have the localized photon flux density, it is therefore necessary to sum $\gamma_{sl}$ over the two emission layers \textit{s}, which can be those of the LED or of the PV cell depending on the device considered. In addition, the volumetric generation rate is needed instead of the global one, and can be written as
\begin{equation}\label{Generation}
    G_{l}=-\frac{d\gamma^{in}}{dz}(z_{l}).
\end{equation}

Similarly, we write the radiative recombination rate as
\begin{equation}\label{Recombination}
    R_{rad}=-\frac{d\gamma^{out}}{dz}(z_{l}).
\end{equation}

Note that in the existing literature for TPV (including in the near-field), radiative recombinations are modeled either by using a radiative recombination coefficient \textit{B} \cite{Francoeur2011,Blandre2017} or by using an iterative process that allows for a precise modelling of photon recycling and transfer between the two bodies \cite{Callahan2021}. Of course, our modelling should tend towards precision (thus the second solution). However, in this work, we use an intermediary between these two models as shown in Eq. \ref{Recombination}, which has several advantages:

\begin{enumerate}
\item compared to the use of a coefficient $B$, using the local emission rate for computing the radiative recombinations allows to account for the influence of the geometry and of near-field effects.
\item compared to the precise modelling, this method does not need any iteration, simplifying the calculations.
\item the expressions of the radiative generation and recombination rates are similarly expressed for the DB and the SDD methods, allowing for an easier comparison.
\end{enumerate}  
The carrier lifetime related to non-radiative recombinations $\tau_{non-rad}$ is computed using classic $A$ and $C$ coefficients (see Table \ref{ParamMod}), taking into account Shockley-Read-Hall (SRH) recombinations and Auger recombinations:
\begin{equation}\label{tauNR}
    \tau_{non-rad}=\left(A+C\cdot N^{2} \right)^{-1}
\end{equation}

Once the charge carrier densities are known, the current density can then be computed as:
\begin{equation}\label{CurrentSDD}
    J=J_n+J_p+J_{dp}
\end{equation}

where $J_n$ and $J_p$ are the minority carrier current densities at the depletion region boundary (whose expression can be obtained from Eq. \ref{DriftDiffusion} by setting $E=0$) and $J_{dp}$ is the current related to the depletion region, which gives (neglecting non-radiative recombinations due to the depletion approximation):
\begin{equation}\label{Jdp}
    J_{dp}=e\int_{dp}\left(G(z)-R_{rad}(z)\right)dz
\end{equation}

The doping levels selected in Table \ref{ParamDevice} have been chosen so that these approximations are valid: by taking as a reference the work from Blandre \textit{et al.} \cite{Blandre2017}, a doping level of $10^{17}$ $\text{cm}^{-3}$ for both the p- and n-regions would be enough to consider that the system is in the low-injection regime. In order to ensure that the approximation is correct, we use larger doping levels: a donor doping level $N_{d}$ of $10^{18}$ $\text{cm}^{-3}$ and an acceptor doping level $N_{a}$ of $10^{20}$ $\text{cm}^{-3}$. Since non-radiative recombinations are neglected in the depletion region (see Eq. \ref{Jdp}), $J_{dp}$ is only approached: therefore, if it becomes a predominant part of the current, this could increase the error made by using the SDD model and require the use of a more precise model. For the geometries considered in this paper (see Table \ref{ParamDevice}), $J_ {dp}$ only accounts for 1 to 10\% of the total current density, because the doping levels are high enough to keep the depletion region thin (of the order of tens of nanometers). We thus expect the SDD model to be precise enough for these different cases.

\section{Analysis of the reference case}\label{ResultsRef}
\subsection{Absorption and Emission}\label{AbsEm}
% Figure 2: system
\begin{figure*}[pos=b!]
    %\adjustbox{valign=t}{\begin{minipage}[t]{.59\textwidth}
        \centering
        \begin{subfigure}[b]{.4\textwidth}
    	    \centering
    		\includegraphics[scale=0.7]{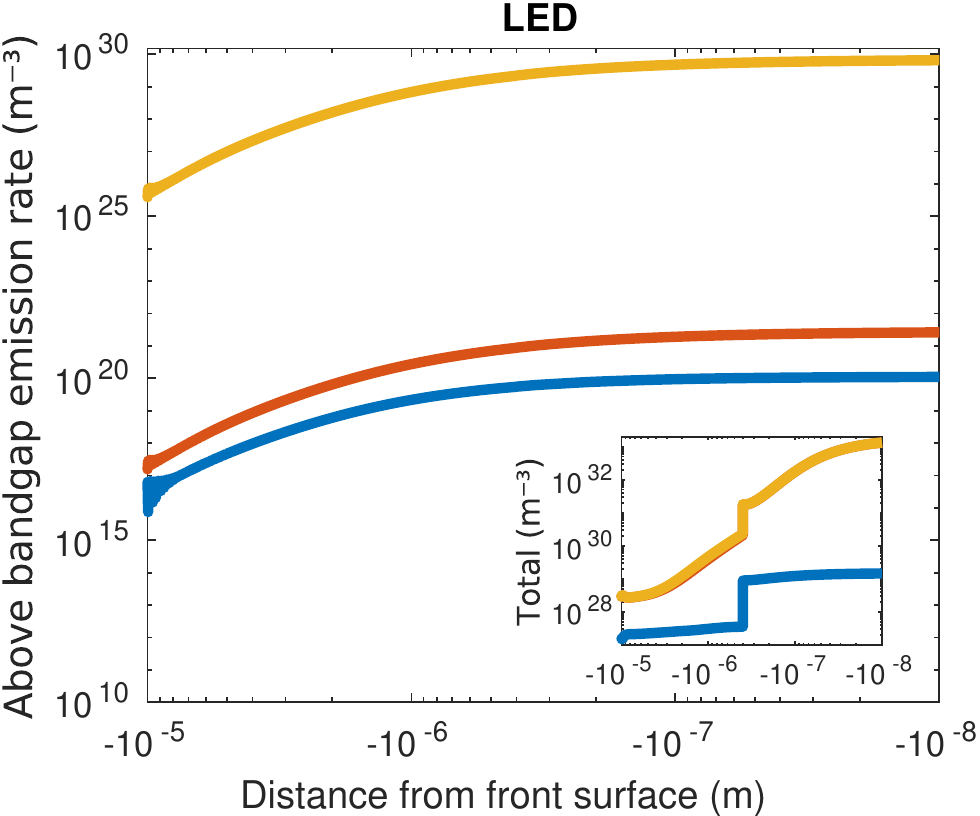}
    	    \caption{}
    	    \label{fig2a}
    	\end{subfigure}
    	\hspace{.3cm}
    	\begin{subfigure}[b]{.4\textwidth}
    	    \centering
    		\includegraphics[scale=0.7]{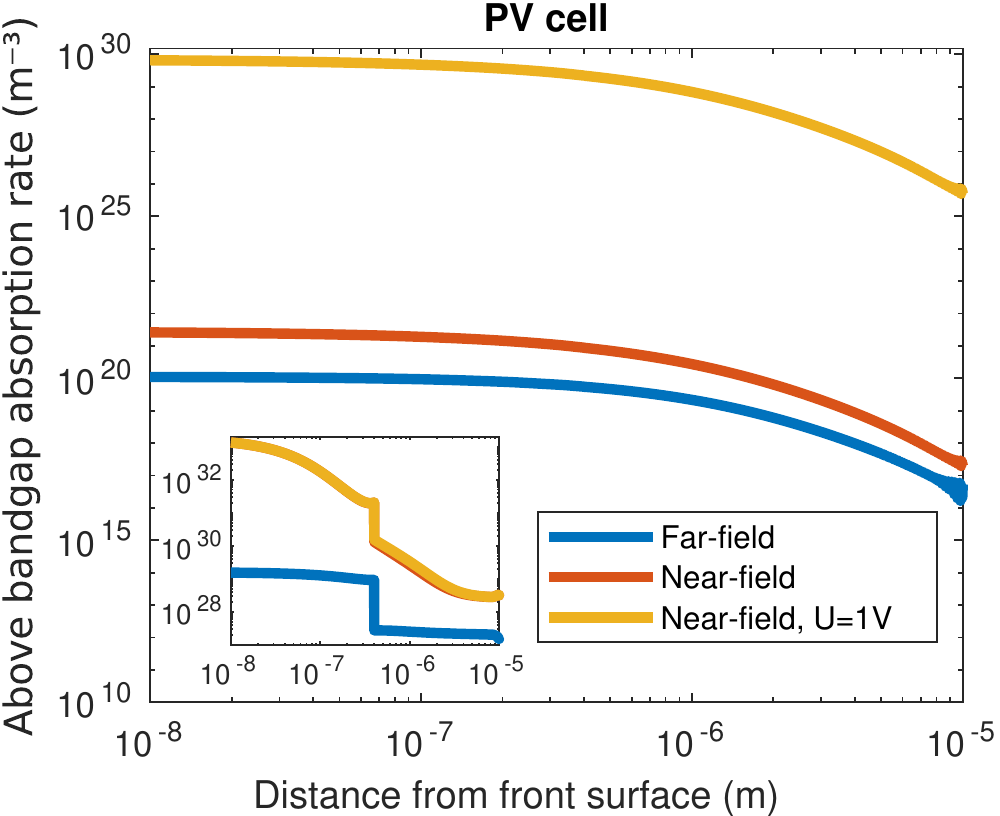}
    	    \caption{}
    	    \label{fig2b}
    	\end{subfigure}
    	\hspace{.6cm}
        
        \hfill
    	\begin{subfigure}[b]{.47\textwidth}
    	    \centering
    		\includegraphics[scale=0.7]{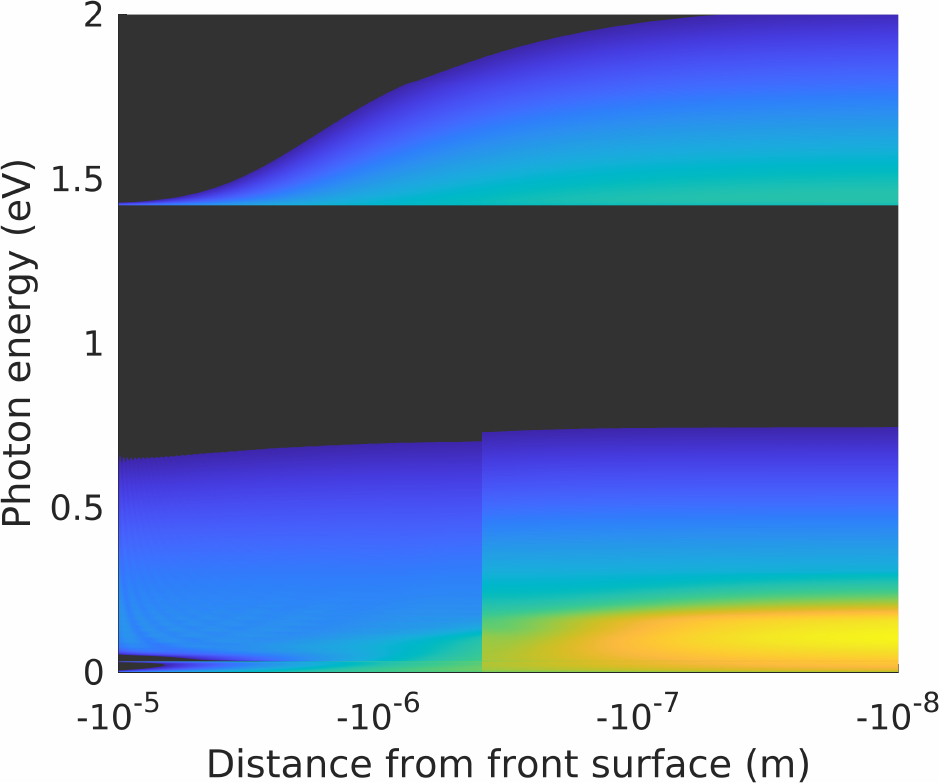}
    	    \caption{}
    	    \label{fig2c}
    	\end{subfigure}
    	\begin{subfigure}[b]{.485\textwidth}
    	    \centering
    		\includegraphics[scale=0.7]{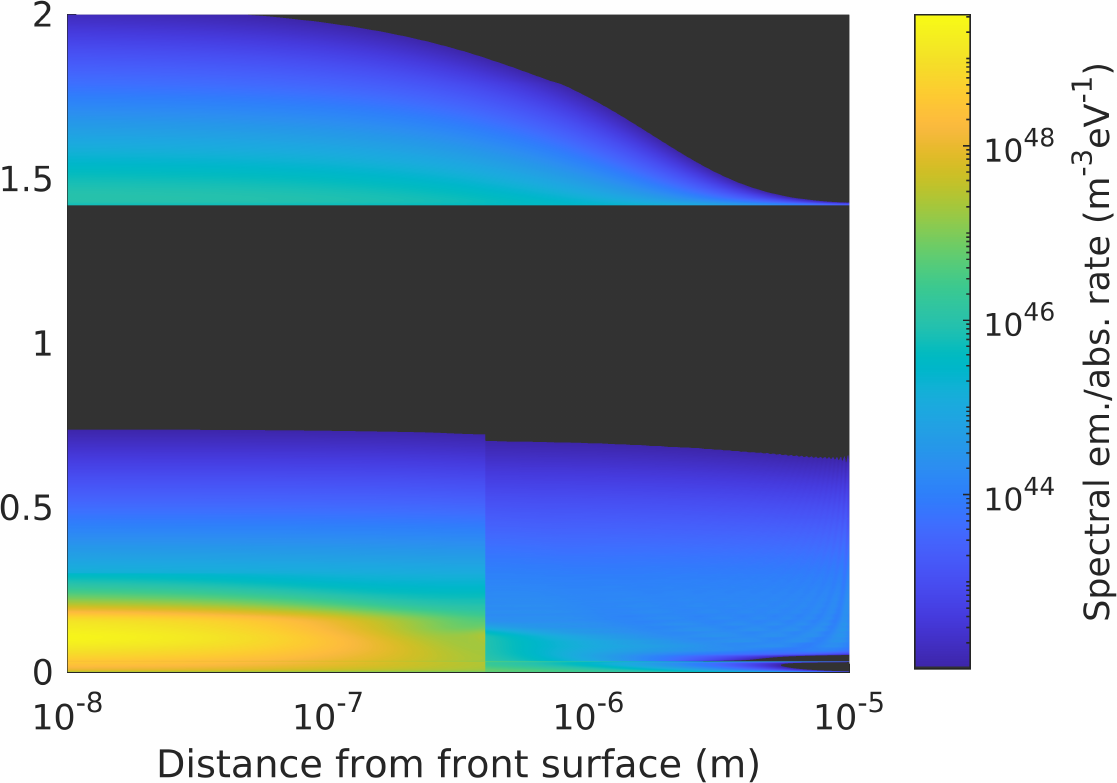}
    	    \caption{}
    	    \label{fig2d}
    	\end{subfigure}
    	
    	\begin{subfigure}[b]{.99\textwidth}
    	    \centering
    		\includegraphics[scale=0.7]{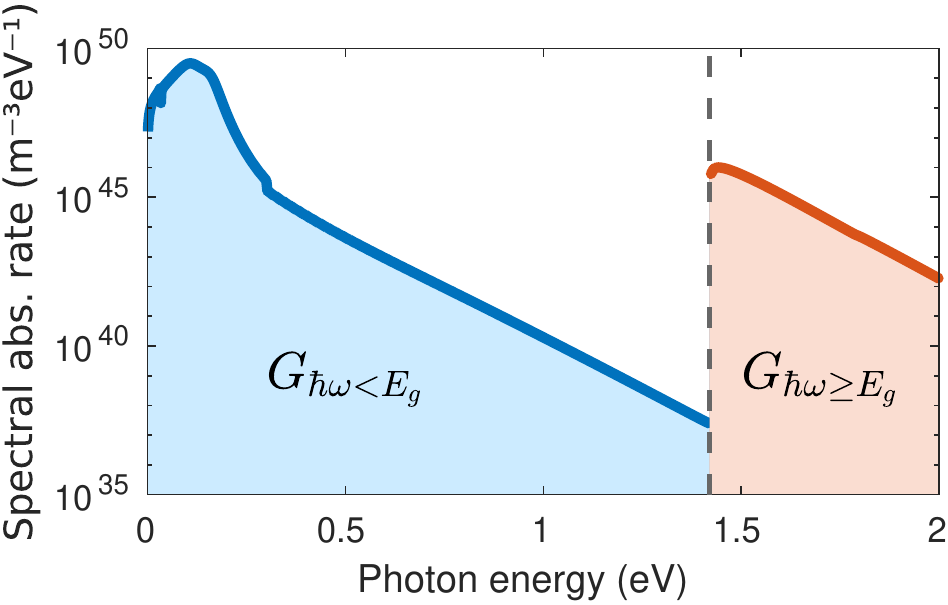}
    	    \caption{}
    	    \label{fig2e}
    	\end{subfigure}
    %\end{minipage}}
    %\adjustbox{valign=t}{\begin{minipage}[t]{.39\textwidth}
        \caption{(a) LED emission profile and (b) PV cell absorption profile of above-bandgap photons in the far-field (blue), in the near-field (orange) and in the near-field with electroluminescence, for LED and PV voltages of 1 V (yellow). Profiles obtained with all photons (below and above bandgap) are shown in insets. The spectral profiles obtained in the near field with electroluminescence are shown for the LED and the PV cell respectively in (c) and (d), for different depths. The corresponding profile at the front surface of the PV cell is shown in (e).}
    	\label{fig2}
    %\end{minipage}}
\end{figure*}
Before looking at the performance of a NF-TPX device, it is interesting to study the net absorption and emission profiles at a 10 nm gap distance, respectively for the LED and the PV cell. The profiles obtained with the reference case (see Table \ref{ParamDevice}) are shown in Figure \ref{fig2}.\newline
Looking first at above-bandgap quantities, which are the most important for power generation, we observe as expected an extremely large increase in the absorption and emission rates, with the mean increase factor from the far-field (FF) case to the near-field (NF) case with electroluminescent enhancement (for LED and PV cell voltages of 1 V) reaching $10^{9}$. Most of this increase is due to electroluminescence: going to the near-field only accounts for a mean increase rate of the order of 10. The dramatic effect of electroluminescence on the radiative heat transfer however comes with an increased electrical power consumption, and does not necessarily cause an increase in the net power density produced. This will be discussed in Section \ref{SecIV}.\newline
In the insets of Figures \ref{fig2a} and \ref{fig2b} are shown the total emission and absorption rates. Their variations are quite different compared to their above-bandgap equivalent. First, the rates abruptly change at the p-n boundary: because the doping levels and the majority carriers are different, the dielectric functions obtained with the Drude model (see Table \ref{ParamMod}) are different for the p and n layers, causing the photons to be reflected back and causing the emission (or absorption) discontinuity.\newline
The rates of increase between the three cases studied are also quite different. Near-field effects allow for a much larger increase, up to $10^4$ at the front surface. Meanwhile, the impact of electroluminescence is nearly negligible, since the orange and yellow curves are almost superimposed. This can be better seen using Figures \ref{fig2c} and \ref{fig2d}, where the spectral emission and absorption rates are shown. Two regions can be distinguished:
\begin{enumerate}[a)]
\item a first one in the infrared (below 0.3 eV, hence above 4 µm), corresponding to the usual photon energy range of radiation for bodies close to room temperature
\item a second one in the visible (between 1.42 and 2 eV, hence between 600 and 900 nm), corresponding to the electrically-induced emission/absorption (i.e., electroluminescence).
\end{enumerate}
As the electroluminescent peak is low compared to the thermal peak (see Figure \ref{fig2e}), we cannot see the influence of the voltage on the total rates.
One can notice the similarity between the LED emission and the PV cell absorption profiles. These two profiles are indeed symmetrical, because of the device symmetry (in terms of geometry, of optical properties and of doping levels).

\subsection{Parametric Study}\label{SensStudy}
\begin{figure}[hb!]
    \centering
    \begin{subfigure}{.42\textwidth}
	    \centering
		\includegraphics[scale=0.7]{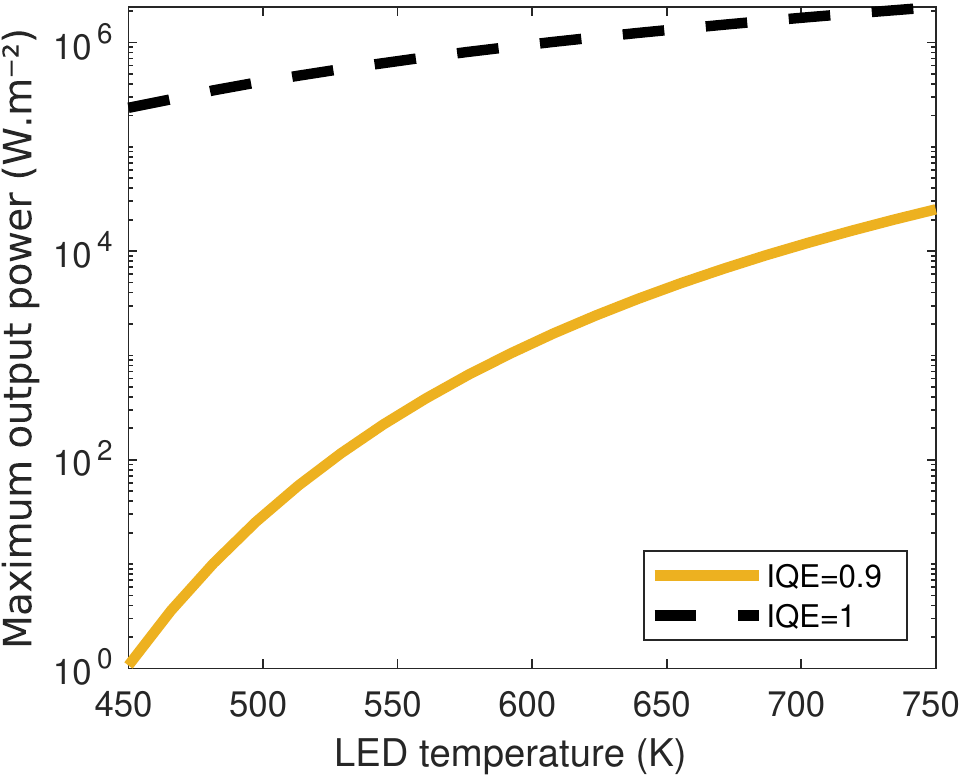}
	    \caption{}
	    \label{figSensStudya}
	\end{subfigure}

    \vspace{.5cm}
	\begin{subfigure}{.42\textwidth}
	    \centering
		\includegraphics[scale=0.7]{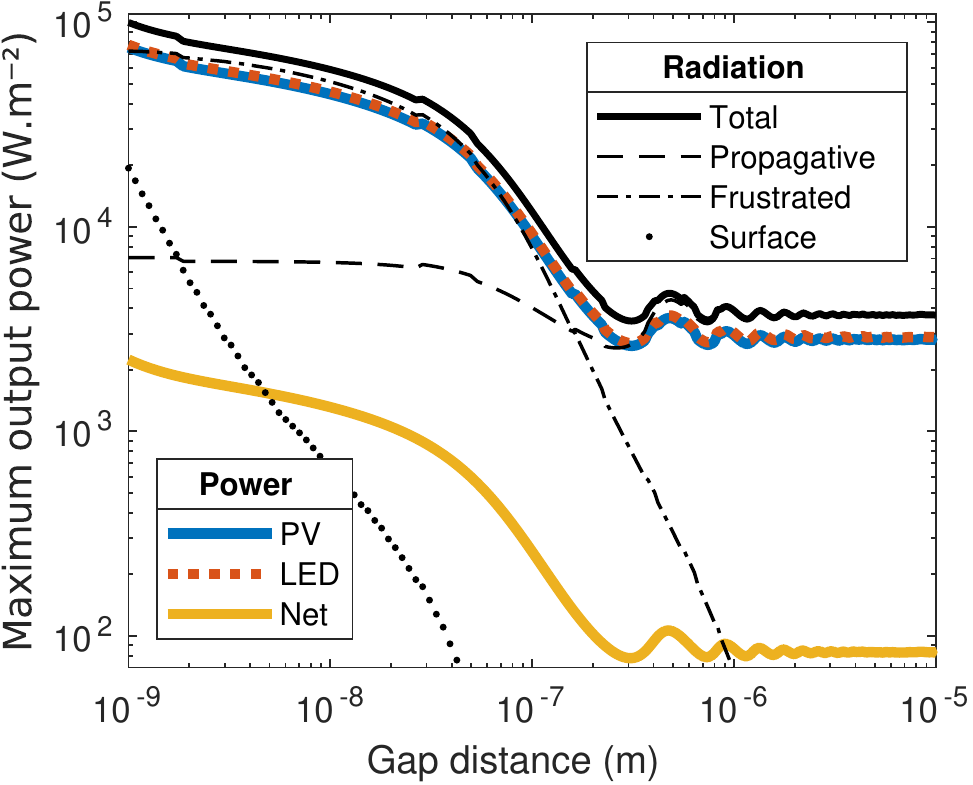}
	    \caption{}
	    \label{figSensStudyb}
	\end{subfigure}
	\caption{Influence of (a) the LED temperature and (b) the gap distance on the power produced by the device. For both cases, the parametric analysis is performed around the reference case. In (b), an IQE of 0.9 is considered.}
	\label{figSensStudy}
\end{figure}
We now investigate the global influence of two key parameters, the LED temperature and the gap distance, on the ideal performance achievable by a GaAs NF-TPX device. In order to do so, we perform a parametric study using the DB method around the reference case (see Table \ref{ParamDevice}). Results are reported in Figure \ref{figSensStudy}. The LED temperature impact on the device is computed for two different IQEs of 1 and 0.9 (frequency and temperature independent). In the ideal scenario ($\text{IQE}=1$), it is shown to have only limited effect on the power output, with an increase rate of a factor 2 between 550 K and 650 K. However, this increase is much larger for lower IQE, reaching a factor 16 in the same range of temperature for $\text{IQE}=0.9$. This shows the importance of using devices with nearly ideal IQE for low temperature application, as stated in the introduction.\newline
In Figure \ref{figSensStudyb} is shown the influence of the gap distance on the above-bandgap radiative heat flux and on the different electrical powers involved, for an IQE of 0.9. As shown in \cite{Bernardi2015}, it is interesting to study the variation of amplitude of the different modes composing the heat flux, i.e. propagative modes, frustrated modes (which propagate in GaAs but are evanescent in vacuum) and surface modes (which are evanescent on each side of the interface). Three different regions can be observed:
\begin{enumerate}[a)]
\item above a few microns: it is the FF regime, with constant values for the different powers.
\item between hundreds of nanometers and a few microns: propagative modes are still dominant in the radiative heat flux, but cavity effects cause oscillations of the powers around the far-field value.
\item below hundreds of nanometers: frustrated evanescent modes become dominant, causing a large increase of the powers at first, before a slower increase below tens of nanometers. At 1 nm, surface modes are still in minority, but have exceeded propagative modes.
\end{enumerate}

Looking more closely at the power levels consumed by the LED and produced by the PV cell, we can note their proximity, caused by an IQE already far from the ideal case and causing the net power output to be quite small compared to them.\newline
By computing the ratio between the radiative heat flux exchanged and the electrical power produced or consumed, we can obtain an estimation of the conversion efficiencies at maximum power. We first focus on above-bandgap photons. Above-bandgap efficiencies are nearly independent of the gap distance, reached at the same voltages for any of the gap distance considered ($U_{LED}=1.00 \text{V}$ and $U_{PV}=1.16 \text{V}$), and equal to
\begin{subequations}\label{ConvEff}
    \begin{gather}
        \eta_{PV,light-el}=\frac{P_{PV}}{q_{\hbar\omega\geq E_{g}}}=78 \%, \\
        \eta_{LED,el-light}=\frac{q_{\hbar\omega\geq E_{g}}}{P_{LED}}=132 \%,
    \end{gather}
\end{subequations}

showing thus the quality of the conversions of each device. Note that the efficiency is greater than 1 for the LED because it is heated: it will still have a positive net emission even with no electrical power consumed. Since most of the power produced by the PV cell is sent back to the LED, the net conversion efficiency is much lower:
\begin{equation}\label{GlobalConvEff}
    \eta_{c}=\frac{P_{net}}{q_{\hbar\omega\geq E_{g}}}=2 \%.
\end{equation}

This net conversion efficiency should not be confused with the efficiency of the device, which would be expressed as the ratio between the device power output and the heat supplied to the LED in order to keep its temperature constant. We obtain
\begin{equation}\label{GlobalEff}
    \eta_{\Sigma}=\frac{P_{net}}{q_{\hbar\omega\geq E_{g}}-P_{LED}}=9 \%.
\end{equation}

equivalent to a scaled efficiency (i.e. the efficiency divided by the related Carnot efficiency) of 18\%, close to what is found for NF-TPV (e.g., 23\% in \cite{Francoeur2011}). These results are obtained for above-bandgap photons, providing an upper bound for the overall efficiency.\newline
A complete evaluation of the efficiency is performed at $d=10$ nm. We model GaAs dielectric function below the bandgap with a decoupled Drude-Lorentz model. Most of the values needed in this model are given or can be computed thanks to \cite{Adachi1994}; for the variation of the Drude damping coefficient, we use the expression given in \cite{Losego2009}. However, since the Drude model depends on the doping levels, we have to consider a specific geometry. By taking the reference case, the junctions are thick enough so that the hypothesis of semi-infinite junctions remains true. The power densities are thus similar to those computed above.\newline
The results obtained are provided in Table \ref{tab:Eff}: as shown in the previous section, below-bandgap photons still dominate the photon flux for an IQE of 0.9, worsening therefore the different efficiencies and increasing the heating (resp. cooling) power required to keep the LED (resp. the PV cell) at constant temperature. While only $10^{4}$ $\text{W}.\text{m}^{-2}$ are needed when considering above-bandgap photons, this rises up to $10^{6}$ $\text{W}.\text{m}^{-2}$ by considering all photons. This underlines the thermal management issue of such devices.
% Table : Device efficiency
\begin{table}[<options>]
\caption{Device efficiency for a 10 nm gap distance.}\label{tab:Eff}
\begin{tabular*}{\tblwidth}{@{}LLLLLL@{}}
\toprule
 & $q$ $(\text{W}.\text{m}^{-2})$ & $\eta_{PV}$ (\%) & $\eta_{LED}$ (\%) & $\eta_c$ (\%) & $\eta_\Sigma$ (\%)  \\ % Table header row
\midrule
Above $E_g$ & 5.9 $\times 10^4$ & 78 & 1.3 $\times 10^2$ & 2.2 & 9.1 \\
All photons & 1.3 $\times 10^6$ & 3.5 & 2.9 $\times 10^3$ & 9.9$\times 10^{-2}$ & 1.0 \\
\bottomrule
\end{tabular*}
\end{table}
\newpage
\subsection{I-V and P-V characteristics}\label{SecIV}
\begin{figure}[htp!]
    \centering
    \hspace{.55cm}
    \begin{subfigure}{.43\textwidth}
	    \centering
		\includegraphics[scale=0.7]{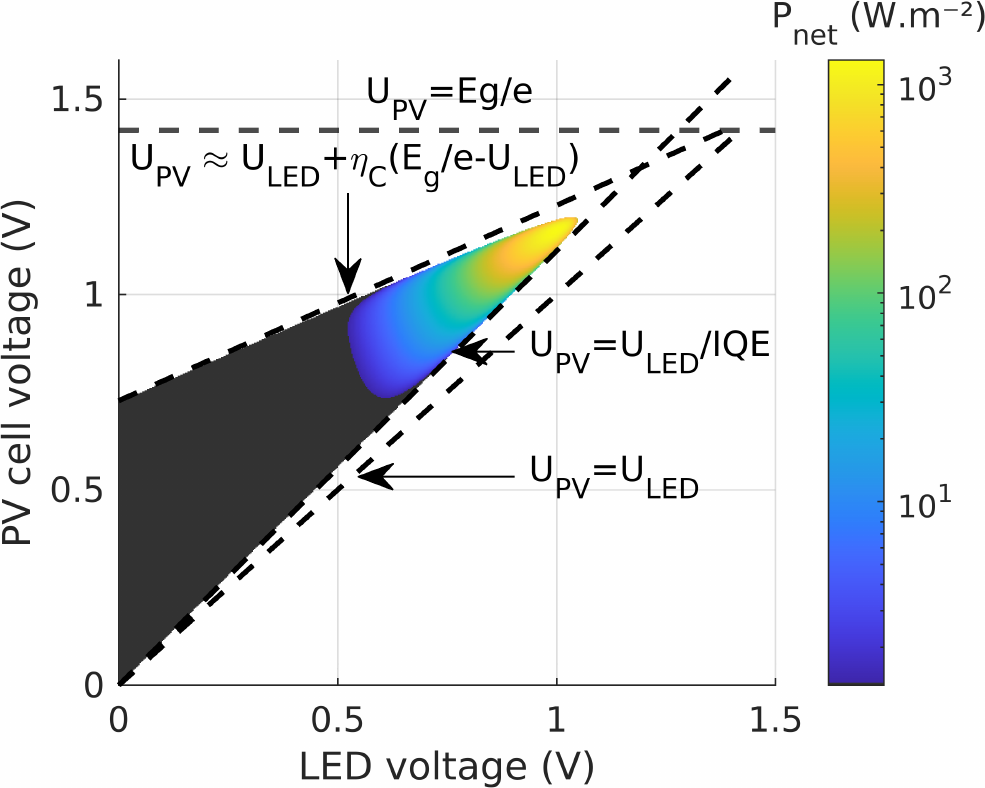}
	    \caption{}
	    \label{FullIV}
	\end{subfigure}
    
    \vspace{.5cm}
	\begin{subfigure}{.4\textwidth}
	    \centering
		\includegraphics[scale=0.7]{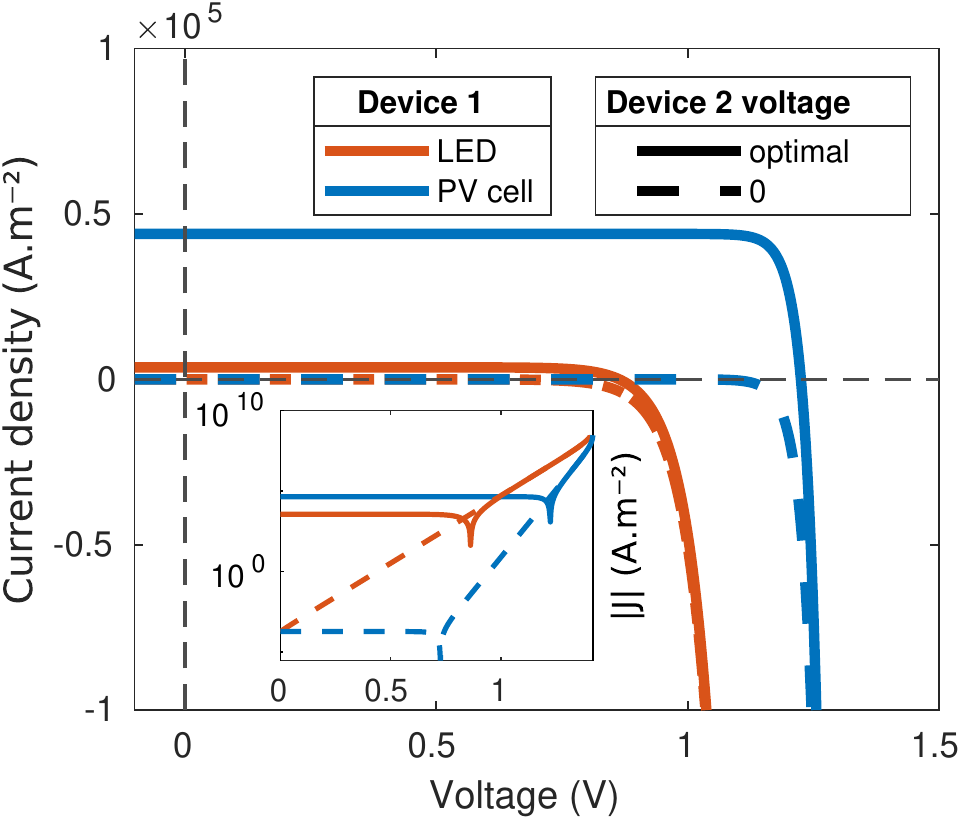}
	    \caption{}
	    \label{IV}
	\end{subfigure}
	
	\vspace{.5cm}
	\begin{subfigure}{.4\textwidth}
	    \centering
		\includegraphics[scale=0.7]{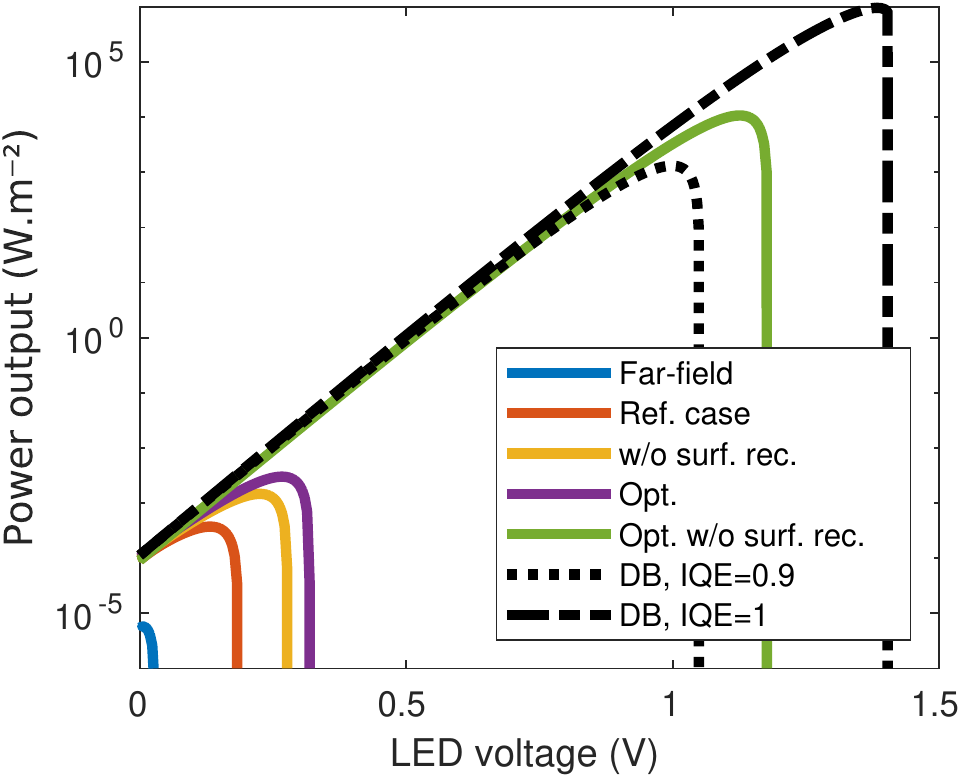}
	    \caption{}
	    \label{SimpIV}
	\end{subfigure}
	\caption{(a) Full V-V characteristic of the device, obtained with the 0D model and for an IQE of 0.9. Precise expression of the upper limit is given in the text. (b) I-V characteristics of the LED and the PV cell around the maximum power point from (a). (c) Simplified P-V characteristics of the device, where the power is computed from the full characteristics as $P(U_{LED})=\underset{U_{PV}}{\mathrm{max}}\:P(U_{LED},U_{PV})$. See Section \ref{ResultsOpti} for an explanation on the optimization.}
	\label{fig4}
\end{figure}

We now look at the characteristics obtained for a NF-TPX device, starting with the results obtained with the DB method for an IQE of 0.9 (Figure \ref{FullIV}). Since both the LED and the PV cell voltages vary, the characteristic of a NF-TPX device is a 3D plot. The region in which the net electrical power output is positive is strictly limited by two inequalities:
\begin{enumerate}[a)]
\item $U_{LED}\leq U_{PV}$: with the hypothesis considered, we always have $|J_{LED}|\geq |J_{PV}|$ (they are equal if $\text{IQE}=1$), having as a consequence that $U_{PV}$ must be greater than $U_{LED}$ in order to produce electrical power.
\item $U_{PV}<E_{g}/e$: if the voltage reaches this value, the modified Bose-Einstein distribution diverges (in reality, this kind of voltage cannot be reached).
\end{enumerate}
Even if the characteristic is limited by these two inequalities, other phenomena determine where power can be extracted from the device. For a given LED voltage, if the PV cell voltage is too large, electroluminescence from the PV cell is important enough to counterbalance the difference of temperature with the LED and change the direction of the net radiative heat flux. In addition, when decreasing the IQE, the range of acceptable PV cell voltage for a given LED voltage becomes narrower, resulting in the decrease of the power output. If the maximum power point (MPP) of an ideal device is close to the point ($E_{g}/e$,$E_{g}/e$), the real MPP is located at lower voltages. A quick estimation of its location can be found at the intersection of two straight lines that qualitatively envelop the characteristic of the device with a good precision and correspond to the two limiting factors stated above:
\begin{enumerate}[a)]
\item $U_{PV}=U_{LED}+\eta_{C}\left(E_{g}/e-U_{LED}\right)+k_BT_{PV}/e\cdot$ $\ln{\left(T_{LED}/T_{PV}\right)}$: balance between thermally- and elec-trically-induced emission for the LED and the PV cell ($\eta_{C}$ corresponds to the Carnot efficiency). This expression is derived from $\gamma=0$, which can be reduced to  $\Delta n^0=0$ by assuming a constant transmission coefficient. As a first approximation, the last term of the equation can be neglected.
\item $U_{PV}=U_{LED}/\text{IQE}$: reduction of the PV cell voltage range available with the IQE.
\end{enumerate}

In Figure \ref{IV}, we show the I-V curve of the LED and the PV cell. The voltage given on the horizontal axis is, respectively, the LED voltage and the PV cell voltage. For the LED (resp. the PV cell) characteristic, the PV cell (resp. the LED) voltage is kept either at its optimum value (i.e. $U_{PV}=1.16$ V and $U_{LED}=1$ V) or at 0. For the PV cell, increasing the LED voltage (from the dashed blue line to the full blue line) raises significantly the short-circuit current by means of electroluminescence. For the LED, we can notice that the variation of the characteristic with the PV cell voltage (from the dashed orange line to the full orange line) is slower; however, for low LED voltage, electroluminescence from the PV cell is large enough to reverse the direction of the net heat flux, thus changing the sign of the short-circuit current and allowing for a power production from the LED. In this case (low LED voltage, high PV cell voltage), the LED works as a PV cell while the PV cell works as a LED. In fact, this regime where the high-temperature device is also the high-voltage device corresponds to the heat-pump/refrigerator configuration of the NF-TPX device.\newline
In order to ease the understanding, we use simplified characteristics in the following, as shown in Figure \ref{SimpIV}. In this case, we plot the variation of the power output only with respect to the LED voltage. For each LED voltage, we take the maximum power output reached for the PV cell voltages considered: $P(U_{LED})=\underset{U_{PV}}{\mathrm{max}}\:P(U_{LED},U_{PV})$. This way of presenting the performance has several advantages:
\begin{enumerate}[a)]
    \item the comparison of the characteristic of different devices is made easier,
    \item the maximum power produced by the TPX device, along with the power produced by the related TPV device (at $U_{LED}=0$), is made more readable.
\end{enumerate}

In black is traced the characteristic obtained with the DB method, for an IQE of 0.9 and 1, with respective maximum output powers of $P_{IQE=0.9}=1.3\times 10^{3}$ $\text{W}.\text{m}^{-2}$ and $P_{IQE=1}=9.7\times 10^{5}$ $\text{W}.\text{m}^{-2}$. Two main domains can be observed, with a fairly linear evolution first, and a sudden decrease in the electrical power output due to the exponential variations caused by the Bose-Einstein distributions. It is also interesting to note that for an IQE of 0.9, the characteristic is nearly identical to the one obtained for $\text{IQE}=1$ up to a LED voltage of 0.9 V, meaning that the reduction of the power output seems to be mainly affected by the reduction of the achievable voltage range.\newline
In color are then plotted the results obtained with the SDD method, in the FF (blue) and in the NF (orange). The maximum performance of the device is given respectively as $P_{FF}=5.8\times 10^{-6}$ $\text{W}.\text{m}^{-2}$ and $P_{NF}=3.7\times 10^{-4}$ $\text{W}.\text{m}^{-2}$. Even in the NF, the performance of the device considered is mediocre. Still, using TPX instead of TPV (i.e., with $U_{LED}=0$ V) shows a quite noteworthy improvement ($P_{NF-TPX}/P_{NF-TPV}\simeq 4$). The different maximum power outputs and their respective LED and PV cell voltages are summarized in Table \ref{TabMPP}.\newline
In the next section, we will thus focus on the understanding of the difference in the results obtained between the DB and the SDD methods, before trying to improve the device considered.
% Table : Modelling parameters
\begin{table}[ht!]
\caption{Maximum power point reached for the different cases studied in Figure \ref{SimpIV}.} \label{TabMPP}
\begin{tabular*}{\tblwidth}{@{}LLLLL@{}}
\toprule
Method & Case & P ($\text{W}.\text{m}^{-2}$) & $\text{U}_{\text{LED}} (\text{V})$ & $\text{U}_{\text{PV}} (\text{V})$  \\
\midrule
SDD & Far-field & 5.8$\times 10^{-6}$ & 0.00 & 0.58 \\
& Ref. case & 3.7$\times 10^{-4}$ & 0.13 & 0.69 \\
& W/o surf. rec. & 1.5$\times 10^{-3}$ & 0.22 & 0.75 \\
& Optimized & 3.0$\times 10^{-3}$ & 0.27 & 0.77 \\
& Opt. w/o surf. rec. & 1.1$\times 10^{4}$ & 1.13 & 1.25 \\
DB & $\text{IQE}=0.9$ & 1.3$\times 10^{3}$ & 1.00 & 1.16 \\
& $\text{IQE}=1$ & 9.7$\times 10^{5}$ & 1.38 & 1.4 \\
\bottomrule
\end{tabular*}
\end{table}

\section{Towards improved devices}\label{ResultsOpti}
\subsection{Surface recombinations}
A first means to increase the performance of the device is to decrease surface recombinations. By nature, these are non-idealities caused by the device (i.e., not necessarily directly related to a given material) that limit the collection of charges. In Figure \ref{SimpIV}, we show the P-V characteristic obtained when surface recombinations are removed (in yellow). If this allows for moderate improvement, with the maximum power output increased by a factor 4 compared to the reference case and reaching $P_{S=0}=1.5\times 10^{-3}$ $\text{W}.\text{m}^{-2}$, reducing the surface recombinations is still important to approach quasi-ideal and ideal cases as will be discussed in Section \ref{ImprovingGeo}.

\subsection{Carrier diffusion length}
\begin{figure}[t!]
    \centering
    \begin{subfigure}{.49\textwidth}
	    \centering
		\includegraphics[scale=0.7]{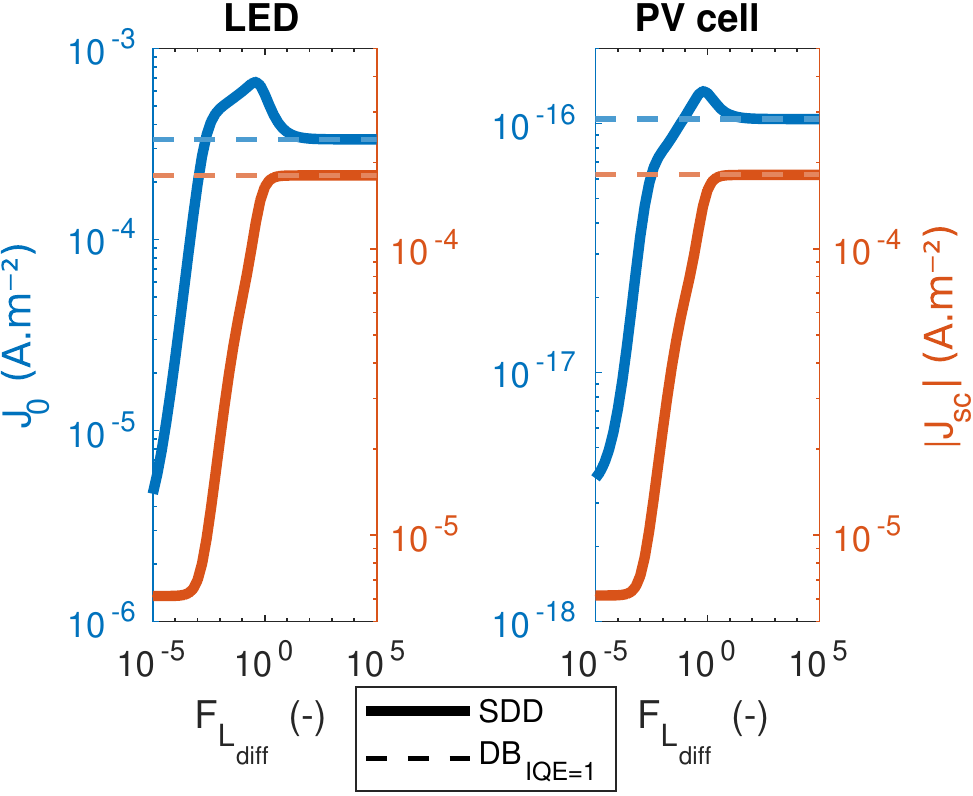}
	\end{subfigure}
	
    \begin{subfigure}{.18\textwidth}
	    \centering
	    \caption{}
	\end{subfigure}
	\begin{subfigure}{.17\textwidth}
	    \centering
	    \caption{}
	\end{subfigure}
	\caption{Impact of the diffusion length $L_{diff}$ on the dark current and the short-circuit current of (a) an LED at 600 K, facing a GaAs absorber at 300 K and (b) a PV cell at 300 K facing a GaAs emitter at 600 K. Since the value of the diffusion length changes in the p-region and in the n-region, the variation of the dark current is shown as a function of the factor $F_{L}$ defined as $L_{diff}=F_{L}L_{diff,0}$. Similar tendencies are found for the short-circuit current.}
	\label{varDtau}
\end{figure}

Two other parameters of interest could be studied in order to search for improved performance: the carriers diffusion coefficient \textit{D} (Eq. \ref{DiffCoeff}) and the non-radiative lifetime $\tau$ (Eq. \ref{tauNR}). Unlike surface recombinations coefficients, these two parameters are physically only related to the material and thus cannot be changed easily. However, they provide good information about the charge carrier movement capabilities inside the device, through the diffusion length $L_{diff}=\sqrt{D\tau}$. By varying $L_{diff}$, we modify the distance carriers will be able to travel, a process which is thus equivalent to changing the geometry for electrons and holes, but not for photons. By doing this theoretical analysis, we can easily observe how close we could approach the ideal case; this provides information about what could be obtained when optimizing the geometry. For the sake of simplicity, this study is performed separately for the LED and the PV cell, with the other device kept as a passive emitter or receiver. We choose to change the diffusion length around the value $L_{diff,0}$ obtained using the physical parameters from Table \ref{ParamMod}. Since both \textit{D} and $\tau$ change when the diffusion length varies, we force them to vary at the same rate: $L_{diff}=\sqrt{(F_L\cdot D_{0})(F_L\cdot \tau_{0})}=F_{L}L_{diff,0}$. The results, shown in Figure \ref{varDtau}, reveal that the I-V characteristic of the two devices tends towards the ideal case when the diffusion lengths considered reach a hundred times the physical value. By a really rough approximation, we could state that increasing the diffusion length by a factor $F_{L}$ is similar to decreasing the geometry by the same factor. While it is probably not possible to decrease all the dimensions by a factor 100, we can however note that the dark current is already quite close to the ideal value for $F_{L}=10$.

\subsection{Improving the geometry}\label{ImprovingGeo}
We now try to optimize the geometry in order to see if we could approach the ideal case. This optimization is performed through the \textit{fmincon} solver of MATLAB, an algorithm searching for local minima; if this solution would not be relevant for finding the optimal geometry, since the result of the process can vary with the starting point, it however returns geometries that are improved in comparison to the reference case.\newline
We set the thickness for each region in the range [10 nm;10 µm], while eliminating surface recombinations. The result of this early optimization process is given in Figure \ref{SimpIV} (green), while the improved geometry parameters are given in Table \ref{ParamDevice}. The maximum power output is much larger than those obtained for all the other cases, reaching a value of $P_{opt,S=0}=1.1\times 10^{4}$ $\text{W}.\text{m}^{-2}$. It even exceeds what was obtained with the DB method for a quasi-ideal NF-TPX device with an IQE of 0.9, and the performance reached by a FF-TPV device with an emitter at more than 1400 K \cite{Fan2020}. It is also close to what is found for NF-TPV for 700 K temperature difference \cite{Vaillon2019}, and for thermoelectric devices for 300 K temperature difference \cite{ElOualid2021}. This result is invigorating for the development of NF-TPX device: it shows that even when getting rid of idealizing approximations, we can reach really high electrical power densities by improving both the contacts at the boundaries of the LED and the PV cell and their geometries for efficient charge carrier management. The device performance could be even further improved by optimizing the doping level of the different layers; however, the use of the SDD method limits the range of doping levels for which the result obtained is accurate. Solving instead the full Drift-Diffusion equations (see Eq. \ref{Poisson},\ref{Continuity} and \ref{DriftDiffusion}) would permit a correct optimization of the doping levels.\newline
\setlength{\textfloatsep}{\baselineskip}
\begin{figure}[b!]
\centering
    \hfill
    \begin{subfigure}{.217\textwidth}
	    \centering
		\includegraphics[scale=0.7]{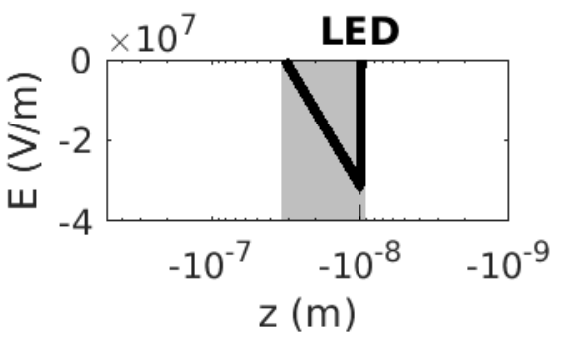}
	    \caption{}
	    \label{fig6a}
	\end{subfigure}
	\begin{subfigure}{.23\textwidth}
	    \centering
		\includegraphics[scale=0.7]{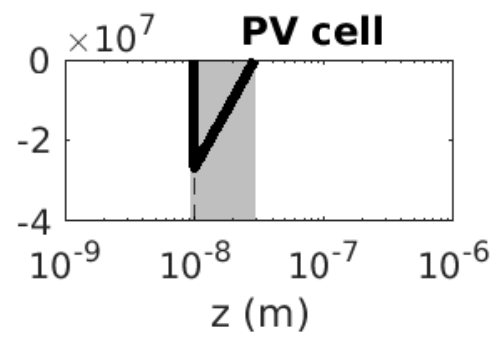}
	    \caption{}
	    \label{fig6b}
	\end{subfigure}
    
    \vspace{.1cm}
    \hfill
    \begin{subfigure}{.225\textwidth}
	    \centering
		\includegraphics[scale=0.7]{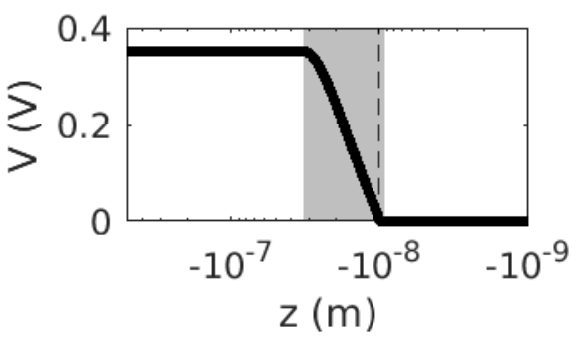}
	    \caption{}
	    \label{fig6c}
	\end{subfigure}
	\begin{subfigure}{.23\textwidth}
	    \centering
		\includegraphics[scale=0.7]{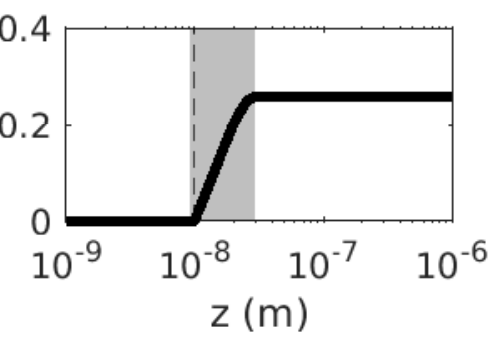}
	    \caption{}
	    \label{fig6d}
	\end{subfigure}
    
    \vspace{.1cm}
    \hfill
	\begin{subfigure}{.23\textwidth}
	    \centering
		\includegraphics[scale=0.7]{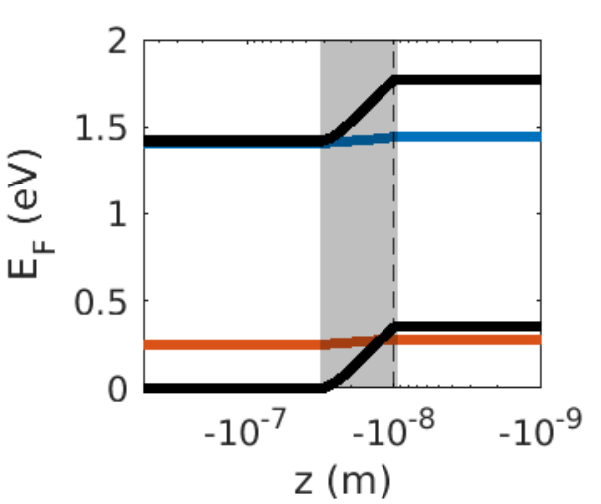}
	    \caption{}
	    \label{fig6e}
	\end{subfigure}
	\begin{subfigure}{.23\textwidth}
	    \centering
		\includegraphics[scale=0.7]{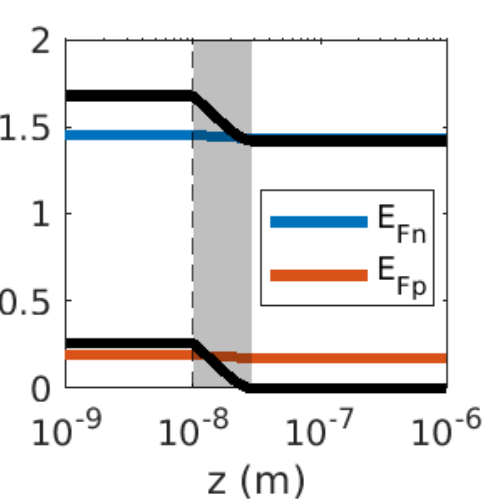}
	    \caption{}
	    \label{fig6f}
	\end{subfigure}
	\caption{(a,b) Electric field, (c,d) Electrostatic potential and (e,f) Quasi-Fermi levels of the LED and the PV cell composing the device at the maximum power point reached without surface recombinations, with the optimized geometry at 10 nm. The depletion region is represented in grey.}
	\label{fig6}
\end{figure}It is important to notice that the improvement of the geometry alone would not allow for such a large increase of the power output: when optimizing the geometry while keeping surface recombinations on the front surface (in purple in Figure \ref{SimpIV}, see Table \ref{ParamDevice} for the geometry obtained), we reach $P_{opt}=3.0\times 10^{-3}$ $\text{W}.\text{m}^{-2}$, in the same order of magnitude as what was found using the reference case without surface recombinations.\newline
The optimized geometry exhibits p-regions that are only 10 nm thin (i.e., the lower bound of the thickness range tested); if reaching such thickness seems difficult to achieve, it highlights nonetheless the necessity to reduce the thickness of the front layer in order to achieve high performance.\newline
In Figure \ref{fig6}, we show the variation of electric field, electrostatic potential and quasi Fermi levels inside the LED and the PV cell, at the MPP. We notice that the quasi-Fermi levels (in color in Figures \ref{fig6e} and \ref{fig6f}) are nearly constant: the assumption that the electrochemical potential is nearly constant and equal to $eU$ seems therefore valid for this specific case.

\section{Conclusion and Outlooks}\label{Ccl}
We have developed a model coupling near-field radiative heat transfer and charge carrier transport using the Simplified Drift-Diffusion method, allowing for an accurate representation of the phenomena occurring in a near-field thermophotonic device. The difference between the results obtained with the Detailed Balance (in 0D) and the Simplified Drift-Diffusion (in 1D) methods underlines the necessity to use the latter for a precise estimation of a device performance. By carrying an extensive analysis of a GaAs-based near-field thermophotonic device, we have also identified that the diffusion length and the surface recombinations have strong impacts on the performance. Through the improvement of the contacts and of the geometry, we are capable of reaching power densities as high as $P=1.1\times 10^{4}$ $\text{W}.\text{m}^{-2}$, i.e. close to 35 $\text{W}.\text{m}^{-2}.\text{K}^{-1}$. This is a promising result for near-field thermophotonic devices, especially since such performance was obtained in spite of the high GaAs bandgap energy compared to thermal energy at 600 K. In addition, the doping concentrations are fixed in this study, thus not optimized.\newline
Of course, the numerical study developed above has still room for improvement. In future work, the model itself should be improved, by getting rid of the approximation $\mu=eU$ in the modified Bose-Einstein distribution (see Eq. \ref{ModifiedBoseEinstein}) and by solving fully the drift-diffusion equations (see Eq. \ref{Poisson},\ref{Continuity},\ref{DriftDiffusion}) instead of the Simplified Drift-Diffusion. These developments have already been addressed for near-field thermophotovoltaic applications respectively in \cite{Callahan2021} and \cite{Blandre2017}, and could be applied to near-field thermophotonics, allowing for an accurate description of the photon recycling and of the charge carrier behaviour in general. Apart from the algorithm itself, the device considered could also be modified to approach more realistic geometries, by considering the use of pin junctions \cite{Blandre2016} and back reflectors \cite{Zhao2018}.
\newline \newline
We thank E. Blandre for the help given, P. Kivisaari, T. Sadi, J. Oksanen from Aalto University for constructive discussion, and E.J. Tervo for clarification.
\newline \newline
We acknowledge the funding of EU H2020 FET Proactive (EIC) programme through project TPX-Power (GA 951976).

\bibliographystyle{cas-model2-names}

\bibliography{cas-refs}

\begin{thebibliography}{51}
\expandafter\ifx\csname natexlab\endcsname\relax\def\natexlab#1{#1}\fi
\providecommand{\url}[1]{\texttt{#1}}
\providecommand{\href}[2]{#2}
\providecommand{\path}[1]{#1}
\providecommand{\DOIprefix}{doi:}
\providecommand{\ArXivprefix}{arXiv:}
\providecommand{\URLprefix}{URL: }
\providecommand{\Pubmedprefix}{pmid:}
\providecommand{\doi}[1]{\href{http://dx.doi.org/#1}{\path{#1}}}
\providecommand{\Pubmed}[1]{\href{pmid:#1}{\path{#1}}}
\providecommand{\bibinfo}[2]{#2}
\ifx\xfnm\relax \def\xfnm[#1]{\unskip,\space#1}\fi
%Type = Article
\bibitem[{Green and Bremner(2016)}]{Green2016}
\bibinfo{author}{Green, M.A.}, \bibinfo{author}{Bremner, S.P.},
  \bibinfo{year}{2016}.
\newblock \bibinfo{title}{{Energy conversion approaches and materials for
  high-efficiency photovoltaics}}.
\newblock \bibinfo{journal}{Nature Materials} \bibinfo{volume}{16},
  \bibinfo{pages}{23--34}.
%Type = Article
\bibitem[{Melnick and Kaviany(2019)}]{Melnick2019}
\bibinfo{author}{Melnick, C.}, \bibinfo{author}{Kaviany, M.},
  \bibinfo{year}{2019}.
\newblock \bibinfo{title}{{From thermoelectricity to phonoelectricity}}.
\newblock \bibinfo{journal}{Applied Physics Reviews} \bibinfo{volume}{6},
  \bibinfo{pages}{021305}.
%Type = Article
\bibitem[{Chen et~al.(2018)Chen, Zhang and Pei}]{Chen2018a}
\bibinfo{author}{Chen, Z.}, \bibinfo{author}{Zhang, X.}, \bibinfo{author}{Pei,
  Y.}, \bibinfo{year}{2018}.
\newblock \bibinfo{title}{{Manipulation of Phonon Transport in
  Thermoelectrics}}.
\newblock \bibinfo{journal}{Advanced Materials} \bibinfo{volume}{30},
  \bibinfo{pages}{1705617}.
%Type = Article
\bibitem[{Mao et~al.(2018)Mao, Liu, Zhou, Zhu, Zhang, Chen and Ren}]{Mao2018}
\bibinfo{author}{Mao, J.}, \bibinfo{author}{Liu, Z.}, \bibinfo{author}{Zhou,
  J.}, \bibinfo{author}{Zhu, H.}, \bibinfo{author}{Zhang, Q.},
  \bibinfo{author}{Chen, G.}, \bibinfo{author}{Ren, Z.}, \bibinfo{year}{2018}.
\newblock \bibinfo{title}{{Advances in thermoelectrics}}.
\newblock \bibinfo{journal}{Advances in Physics} \bibinfo{volume}{67},
  \bibinfo{pages}{69--147}.
%Type = Article
\bibitem[{Daneshvar et~al.(2015)Daneshvar, Prinja and Kherani}]{Daneshvar2015}
\bibinfo{author}{Daneshvar, H.}, \bibinfo{author}{Prinja, R.},
  \bibinfo{author}{Kherani, N.P.}, \bibinfo{year}{2015}.
\newblock \bibinfo{title}{{Thermophotovoltaics: Fundamentals, challenges and
  prospects}}.
\newblock \bibinfo{journal}{Applied Energy} \bibinfo{volume}{159},
  \bibinfo{pages}{560--575}.
%Type = Article
\bibitem[{Schwede et~al.(2010)Schwede, Bargatin, Riley, Hardin, Rosenthal, Sun,
  Schmitt, Pianetta, Howe, Shen and Melosh}]{Schwede2010}
\bibinfo{author}{Schwede, J.W.}, \bibinfo{author}{Bargatin, I.},
  \bibinfo{author}{Riley, D.C.}, \bibinfo{author}{Hardin, B.E.},
  \bibinfo{author}{Rosenthal, S.J.}, \bibinfo{author}{Sun, Y.},
  \bibinfo{author}{Schmitt, F.}, \bibinfo{author}{Pianetta, P.},
  \bibinfo{author}{Howe, R.T.}, \bibinfo{author}{Shen, Z.X.},
  \bibinfo{author}{Melosh, N.A.}, \bibinfo{year}{2010}.
\newblock \bibinfo{title}{{Photon-enhanced thermionic emission for solar
  concentrator systems}}.
\newblock \bibinfo{journal}{Nature Materials} \bibinfo{volume}{9},
  \bibinfo{pages}{762--767}.
%Type = Article
\bibitem[{Campbell et~al.(2021)Campbell, Celenza, Schmitt, Schwede and
  Bargatin}]{Campbell2021}
\bibinfo{author}{Campbell, M.F.}, \bibinfo{author}{Celenza, T.J.},
  \bibinfo{author}{Schmitt, F.}, \bibinfo{author}{Schwede, J.W.},
  \bibinfo{author}{Bargatin, I.}, \bibinfo{year}{2021}.
\newblock \bibinfo{title}{{Progress Toward High Power Output in Thermionic
  Energy Converters}}.
\newblock \bibinfo{journal}{Advanced Science} \bibinfo{volume}{8},
  \bibinfo{pages}{2003812}.
%Type = Article
\bibitem[{K{\"{o}}nig et~al.(2010)K{\"{o}}nig, Casalenuovo, Takeda, Conibeer,
  Guillemoles, Patterson, Huang and Green}]{Konig2010}
\bibinfo{author}{K{\"{o}}nig, D.}, \bibinfo{author}{Casalenuovo, K.},
  \bibinfo{author}{Takeda, Y.}, \bibinfo{author}{Conibeer, G.},
  \bibinfo{author}{Guillemoles, J.F.}, \bibinfo{author}{Patterson, R.},
  \bibinfo{author}{Huang, L.M.}, \bibinfo{author}{Green, M.A.},
  \bibinfo{year}{2010}.
\newblock \bibinfo{title}{{Hot carrier solar cells: Principles, materials and
  design}}.
\newblock \bibinfo{journal}{Physica E: Low-Dimensional Systems and
  Nanostructures} \bibinfo{volume}{42}, \bibinfo{pages}{2862--2866}.
%Type = Article
\bibitem[{Li et~al.(2019)Li, Fu, Xu and Sum}]{Li2019}
\bibinfo{author}{Li, M.}, \bibinfo{author}{Fu, J.}, \bibinfo{author}{Xu, Q.},
  \bibinfo{author}{Sum, T.C.}, \bibinfo{year}{2019}.
\newblock \bibinfo{title}{{Slow Hot-Carrier Cooling in Halide Perovskites:
  Prospects for Hot-Carrier Solar Cells}}.
\newblock \bibinfo{journal}{Advanced Materials} \bibinfo{volume}{31},
  \bibinfo{pages}{1802486}.
%Type = Article
\bibitem[{Sakakibara et~al.(2019)Sakakibara, Stelmakh, Chan, Ghebrebrhan,
  Joannopoulos, Solja{\v{c}}i{\'{c}} and
  {\v{C}}elanovi{\'{c}}}]{Sakakibara2019}
\bibinfo{author}{Sakakibara, R.}, \bibinfo{author}{Stelmakh, V.},
  \bibinfo{author}{Chan, W.R.}, \bibinfo{author}{Ghebrebrhan, M.},
  \bibinfo{author}{Joannopoulos, J.D.}, \bibinfo{author}{Solja{\v{c}}i{\'{c}},
  M.}, \bibinfo{author}{{\v{C}}elanovi{\'{c}}, I.}, \bibinfo{year}{2019}.
\newblock \bibinfo{title}{{Practical emitters for thermophotovoltaics: a
  review}}.
\newblock \bibinfo{journal}{Journal of Photonics for Energy}
  \bibinfo{volume}{9}, \bibinfo{pages}{032713}.
%Type = Article
\bibitem[{Fan et~al.(2020)Fan, Burger, McSherry, Lee, Lenert and
  Forrest}]{Fan2020}
\bibinfo{author}{Fan, D.}, \bibinfo{author}{Burger, T.},
  \bibinfo{author}{McSherry, S.}, \bibinfo{author}{Lee, B.},
  \bibinfo{author}{Lenert, A.}, \bibinfo{author}{Forrest, S.R.},
  \bibinfo{year}{2020}.
\newblock \bibinfo{title}{{Near-perfect photon utilization in an air-bridge
  thermophotovoltaic cell}}.
\newblock \bibinfo{journal}{Nature} \bibinfo{volume}{586},
  \bibinfo{pages}{237--241}.
%Type = Article
\bibitem[{Tervo et~al.(2018)Tervo, Bagherisereshki and Zhang}]{Tervo2018}
\bibinfo{author}{Tervo, E.J.}, \bibinfo{author}{Bagherisereshki, E.},
  \bibinfo{author}{Zhang, Z.M.}, \bibinfo{year}{2018}.
\newblock \bibinfo{title}{{Near-field radiative thermoelectric energy
  converters: a review}}.
\newblock \bibinfo{journal}{Frontiers in Energy} \bibinfo{volume}{12},
  \bibinfo{pages}{5--21}.
%Type = Article
\bibitem[{Datas(2016)}]{Datas2016}
\bibinfo{author}{Datas, A.}, \bibinfo{year}{2016}.
\newblock \bibinfo{title}{{Hybrid thermionic-photovoltaic converter}}.
\newblock \bibinfo{journal}{Applied Physics Letters} \bibinfo{volume}{108},
  \bibinfo{pages}{143503}.
%Type = Article
\bibitem[{Zeneli et~al.(2020)Zeneli, Bellucci, Sabbatella, Trucchi,
  Nikolopoulos, Nikolopoulos, Karellas and Kakaras}]{Zeneli2020}
\bibinfo{author}{Zeneli, M.}, \bibinfo{author}{Bellucci, A.},
  \bibinfo{author}{Sabbatella, G.}, \bibinfo{author}{Trucchi, D.M.},
  \bibinfo{author}{Nikolopoulos, A.}, \bibinfo{author}{Nikolopoulos, N.},
  \bibinfo{author}{Karellas, S.}, \bibinfo{author}{Kakaras, E.},
  \bibinfo{year}{2020}.
\newblock \bibinfo{title}{{Performance evaluation and optimization of the
  cooling system of a hybrid thermionic-photovoltaic converter}}.
\newblock \bibinfo{journal}{Energy Conversion and Management}
  \bibinfo{volume}{210}, \bibinfo{pages}{112717}.
%Type = Article
\bibitem[{Whale and Cravalho(2002)}]{Whale2002}
\bibinfo{author}{Whale, M.D.}, \bibinfo{author}{Cravalho, E.G.},
  \bibinfo{year}{2002}.
\newblock \bibinfo{title}{{Modeling and performance of microscale
  thermophotovoltaic energy conversion devices}}.
\newblock \bibinfo{journal}{IEEE Transactions on Energy Conversion}
  \bibinfo{volume}{17}, \bibinfo{pages}{130--142}.
%Type = Article
\bibitem[{Park et~al.(2008)Park, Basu, King and Zhang}]{Park2008}
\bibinfo{author}{Park, K.}, \bibinfo{author}{Basu, S.}, \bibinfo{author}{King,
  W.P.}, \bibinfo{author}{Zhang, Z.M.}, \bibinfo{year}{2008}.
\newblock \bibinfo{title}{{Performance analysis of near-field
  thermophotovoltaic devices considering absorption distribution}}.
\newblock \bibinfo{journal}{Journal of Quantitative Spectroscopy and Radiative
  Transfer} \bibinfo{volume}{109}, \bibinfo{pages}{305--316}.
%Type = Article
\bibitem[{Francoeur et~al.(2011)Francoeur, Vaillon and Meng}]{Francoeur2011}
\bibinfo{author}{Francoeur, M.}, \bibinfo{author}{Vaillon, R.},
  \bibinfo{author}{Meng, M.P.}, \bibinfo{year}{2011}.
\newblock \bibinfo{title}{{Thermal impacts on the performance of nanoscale-gap
  thermophotovoltaic power generators}}.
\newblock \bibinfo{journal}{IEEE Transactions on Energy Conversion}
  \bibinfo{volume}{26}, \bibinfo{pages}{686--698}.
%Type = Article
\bibitem[{Harder and Green(2003)}]{Harder2003}
\bibinfo{author}{Harder, N.P.}, \bibinfo{author}{Green, M.A.},
  \bibinfo{year}{2003}.
\newblock \bibinfo{title}{{Thermophotonics}}.
\newblock \bibinfo{journal}{Semiconductor Science and Technology}
  \bibinfo{volume}{18}, \bibinfo{pages}{S270--S278}.
%Type = Article
\bibitem[{Zhao et~al.(2019)Zhao, Buddhiraju, Santhanam, Chen and
  Fan}]{Zhao2019}
\bibinfo{author}{Zhao, B.}, \bibinfo{author}{Buddhiraju, S.},
  \bibinfo{author}{Santhanam, P.}, \bibinfo{author}{Chen, K.},
  \bibinfo{author}{Fan, S.}, \bibinfo{year}{2019}.
\newblock \bibinfo{title}{{Self-sustaining thermophotonic circuits}}.
\newblock \bibinfo{journal}{Proceedings of the National Academy of Sciences of
  the United States of America} \bibinfo{volume}{116},
  \bibinfo{pages}{11596--11601}.
%Type = Article
\bibitem[{Zhao et~al.(2018)Zhao, Santhanam, Chen, Buddhiraju and
  Fan}]{Zhao2018}
\bibinfo{author}{Zhao, B.}, \bibinfo{author}{Santhanam, P.},
  \bibinfo{author}{Chen, K.}, \bibinfo{author}{Buddhiraju, S.},
  \bibinfo{author}{Fan, S.}, \bibinfo{year}{2018}.
\newblock \bibinfo{title}{{Near-Field Thermophotonic Systems for Low-Grade
  Waste-Heat Recovery}}.
\newblock \bibinfo{journal}{Nano Letters} \bibinfo{volume}{18},
  \bibinfo{pages}{5224--5230}.
%Type = Book
\bibitem[{Zhang(2007)}]{Zhang2007}
\bibinfo{author}{Zhang, Z.M.}, \bibinfo{year}{2007}.
\newblock \bibinfo{title}{{Nano/Microscale heat transfer}}.
\newblock \bibinfo{publisher}{McGraw Hill}.
%Type = Article
\bibitem[{Fiorino et~al.(2018)Fiorino, Zhu, Thompson, Mittapally, Reddy and
  Meyhofer}]{Fiorino2018}
\bibinfo{author}{Fiorino, A.}, \bibinfo{author}{Zhu, L.},
  \bibinfo{author}{Thompson, D.}, \bibinfo{author}{Mittapally, R.},
  \bibinfo{author}{Reddy, P.}, \bibinfo{author}{Meyhofer, E.},
  \bibinfo{year}{2018}.
\newblock \bibinfo{title}{{Nanogap near-field thermophotovoltaics}}.
\newblock \bibinfo{journal}{Nature Nanotechnology} \bibinfo{volume}{13},
  \bibinfo{pages}{806--811}.
%Type = Article
\bibitem[{Inoue et~al.(2019)Inoue, Koyama, Kang, Ikeda, Asano and
  Noda}]{Inoue2019}
\bibinfo{author}{Inoue, T.}, \bibinfo{author}{Koyama, T.},
  \bibinfo{author}{Kang, D.D.}, \bibinfo{author}{Ikeda, K.},
  \bibinfo{author}{Asano, T.}, \bibinfo{author}{Noda, S.},
  \bibinfo{year}{2019}.
\newblock \bibinfo{title}{{One-Chip Near-Field Thermophotovoltaic Device
  Integrating a Thin-Film Thermal Emitter and Photovoltaic Cell}}.
\newblock \bibinfo{journal}{Nano Letters} \bibinfo{volume}{19},
  \bibinfo{pages}{3948--3952}.
%Type = Article
\bibitem[{Bhatt et~al.(2020)Bhatt, Zhao, Roberts, Datta, Mohanty, Lin,
  Hartmann, St-Gelais, Fan and Lipson}]{Bhatt2020}
\bibinfo{author}{Bhatt, G.R.}, \bibinfo{author}{Zhao, B.},
  \bibinfo{author}{Roberts, S.}, \bibinfo{author}{Datta, I.},
  \bibinfo{author}{Mohanty, A.}, \bibinfo{author}{Lin, T.},
  \bibinfo{author}{Hartmann, J.M.}, \bibinfo{author}{St-Gelais, R.},
  \bibinfo{author}{Fan, S.}, \bibinfo{author}{Lipson, M.},
  \bibinfo{year}{2020}.
\newblock \bibinfo{title}{{Integrated near-field thermo-photovoltaics for heat
  recycling}}.
\newblock \bibinfo{journal}{Nature Communications} \bibinfo{volume}{11},
  \bibinfo{pages}{2545}.
%Type = Article
\bibitem[{Lucchesi et~al.(2021)Lucchesi, Cakiroglu, Perez, Taliercio,
  Tourni{\'{e}}, Chapuis and Vaillon}]{Lucchesi2021}
\bibinfo{author}{Lucchesi, C.}, \bibinfo{author}{Cakiroglu, D.},
  \bibinfo{author}{Perez, J.P.}, \bibinfo{author}{Taliercio, T.},
  \bibinfo{author}{Tourni{\'{e}}, E.}, \bibinfo{author}{Chapuis, P.O.},
  \bibinfo{author}{Vaillon, R.}, \bibinfo{year}{2021}.
\newblock \bibinfo{title}{{Near-Field Thermophotovoltaic Conversion with High
  Electrical Power Density and Cell Efficiency above 14{\%}}}.
\newblock \bibinfo{journal}{Nano Letters} \bibinfo{volume}{11},
  \bibinfo{pages}{4524--4529}.
%Type = Article
\bibitem[{Mittapally et~al.(2021)Mittapally, Lee, Zhu, Reihani, Lim, Fan,
  Forrest, Reddy and Meyhofer}]{Mittapally2021}
\bibinfo{author}{Mittapally, R.}, \bibinfo{author}{Lee, B.},
  \bibinfo{author}{Zhu, L.}, \bibinfo{author}{Reihani, A.},
  \bibinfo{author}{Lim, J.W.}, \bibinfo{author}{Fan, D.},
  \bibinfo{author}{Forrest, S.R.}, \bibinfo{author}{Reddy, P.},
  \bibinfo{author}{Meyhofer, E.}, \bibinfo{year}{2021}.
\newblock \bibinfo{title}{{Near-field thermophotovoltaics for efficient heat to
  electricity conversion at high power density}}.
\newblock \bibinfo{journal}{Nature Communications} \bibinfo{volume}{12},
  \bibinfo{pages}{4364}.
%Type = Article
\bibitem[{Inoue et~al.(2021)Inoue, Ikeda, Song, Suzuki, Ishino, Asano and
  Noda}]{Inoue2021a}
\bibinfo{author}{Inoue, T.}, \bibinfo{author}{Ikeda, K.},
  \bibinfo{author}{Song, B.}, \bibinfo{author}{Suzuki, T.},
  \bibinfo{author}{Ishino, K.}, \bibinfo{author}{Asano, T.},
  \bibinfo{author}{Noda, S.}, \bibinfo{year}{2021}.
\newblock \bibinfo{title}{{Integrated Near-Field Thermophotovoltaic Device
  Overcoming Blackbody Limit}}.
\newblock \bibinfo{journal}{ACS Photonics} \bibinfo{volume}{8},
  \bibinfo{pages}{2466--2472}.
%Type = Article
\bibitem[{Oksanen and Tulkki(2010)}]{Oksanen2010}
\bibinfo{author}{Oksanen, J.}, \bibinfo{author}{Tulkki, J.},
  \bibinfo{year}{2010}.
\newblock \bibinfo{title}{{Thermophotonic heat pump-a theoretical model and
  numerical simulations}}.
\newblock \bibinfo{journal}{Journal of Applied Physics} \bibinfo{volume}{107},
  \bibinfo{pages}{093106}.
%Type = Article
\bibitem[{Sadi et~al.(2019)Sadi, Radevici, Kivisaari and Oksanen}]{Sadi2019a}
\bibinfo{author}{Sadi, T.}, \bibinfo{author}{Radevici, I.},
  \bibinfo{author}{Kivisaari, P.}, \bibinfo{author}{Oksanen, J.},
  \bibinfo{year}{2019}.
\newblock \bibinfo{title}{{Electroluminescent Cooling in III-V Intracavity
  Diodes: Efficiency Bottlenecks}}.
\newblock \bibinfo{journal}{IEEE Transactions on Electron Devices}
  \bibinfo{volume}{66}, \bibinfo{pages}{2651--2656}.
%Type = Article
\bibitem[{Sadi et~al.(2020)Sadi, Radevici and Oksanen}]{Sadi2020}
\bibinfo{author}{Sadi, T.}, \bibinfo{author}{Radevici, I.},
  \bibinfo{author}{Oksanen, J.}, \bibinfo{year}{2020}.
\newblock \bibinfo{title}{{Thermophotonic cooling with light-emitting diodes}}.
\newblock \bibinfo{journal}{Nature Photonics} \bibinfo{volume}{14},
  \bibinfo{pages}{205--214}.
%Type = Article
\bibitem[{Bender et~al.(2013)Bender, Cederberg, Wang and
  Sheik-Bahae}]{Bender2013a}
\bibinfo{author}{Bender, D.A.}, \bibinfo{author}{Cederberg, J.G.},
  \bibinfo{author}{Wang, C.}, \bibinfo{author}{Sheik-Bahae, M.},
  \bibinfo{year}{2013}.
\newblock \bibinfo{title}{{Development of high quantum efficiency GaAs/GaInP
  double heterostructures for laser cooling}}.
\newblock \bibinfo{journal}{Applied Physics Letters} \bibinfo{volume}{102},
  \bibinfo{pages}{252102}.
%Type = Article
\bibitem[{Vaillon et~al.(2020)Vaillon, Parola, Lamnatou and
  Chemisana}]{Vaillon2020}
\bibinfo{author}{Vaillon, R.}, \bibinfo{author}{Parola, S.},
  \bibinfo{author}{Lamnatou, C.}, \bibinfo{author}{Chemisana, D.},
  \bibinfo{year}{2020}.
\newblock \bibinfo{title}{{Solar Cells Operating under Thermal Stress}}.
\newblock \bibinfo{journal}{Cell Reports Physical Science} \bibinfo{volume}{1},
  \bibinfo{pages}{100267}.
%Type = Book
\bibitem[{Nelson(2003)}]{Nelson2003}
\bibinfo{author}{Nelson, J.}, \bibinfo{year}{2003}.
\newblock \bibinfo{title}{{The Physics of Solar Cells}}.
\newblock \bibinfo{publisher}{Imperial College Press}.
%Type = Book
\bibitem[{Kitai(2011)}]{Kitai2011}
\bibinfo{author}{Kitai, A.}, \bibinfo{year}{2011}.
\newblock \bibinfo{title}{{Principles of Solar Cells, LEDs and Diodes: The role
  of the PN junction}}.
\newblock \bibinfo{publisher}{Wiley}.
%Type = Article
\bibitem[{Maros et~al.(2015)Maros, Gangam, Fang, Smith, Vasileska, Goodnick,
  Bertoni and Honsberg}]{Maros2015}
\bibinfo{author}{Maros, A.}, \bibinfo{author}{Gangam, S.},
  \bibinfo{author}{Fang, Y.}, \bibinfo{author}{Smith, J.},
  \bibinfo{author}{Vasileska, D.}, \bibinfo{author}{Goodnick, S.M.},
  \bibinfo{author}{Bertoni, M.I.}, \bibinfo{author}{Honsberg, C.B.},
  \bibinfo{year}{2015}.
\newblock \bibinfo{title}{{High temperature characterization of GaAs single
  junction solar cells}}.
\newblock \bibinfo{journal}{2015 IEEE 42nd Photovoltaic Specialist Conference,
  PVSC 2015} .
%Type = Article
\bibitem[{Sun et~al.(2017)Sun, Faucher, Jung, Vaisman, McPheeters, Sharps,
  Perl, Simon, Steiner, Friedman and Lee}]{Sun2017}
\bibinfo{author}{Sun, Y.}, \bibinfo{author}{Faucher, J.},
  \bibinfo{author}{Jung, D.}, \bibinfo{author}{Vaisman, M.},
  \bibinfo{author}{McPheeters, C.}, \bibinfo{author}{Sharps, P.},
  \bibinfo{author}{Perl, E.}, \bibinfo{author}{Simon, J.},
  \bibinfo{author}{Steiner, M.}, \bibinfo{author}{Friedman, D.J.},
  \bibinfo{author}{Lee, M.L.}, \bibinfo{year}{2017}.
\newblock \bibinfo{title}{{Thermal stability of GaAs solar cells for high
  temperature applications}}.
\newblock \bibinfo{journal}{2017 IEEE 44th Photovoltaic Specialist Conference
  (PVSC)} , \bibinfo{pages}{2385--2388}.
%Type = Article
\bibitem[{Gonzalez-Cuevas et~al.(2007)Gonzalez-Cuevas, Refaat, Abedin and
  Elsayed-Ali}]{Gonzalez-Cuevas2007}
\bibinfo{author}{Gonzalez-Cuevas, J.A.}, \bibinfo{author}{Refaat, T.F.},
  \bibinfo{author}{Abedin, M.N.}, \bibinfo{author}{Elsayed-Ali, H.E.},
  \bibinfo{year}{2007}.
\newblock \bibinfo{title}{{Calculations of the temperature and alloy
  composition effects on the optical properties of Alx Ga1-x Asy Sb1-y and Gax
  In1-x Asy Sb1-y in the spectral range 0.5-6 eV}}.
\newblock \bibinfo{journal}{Journal of Applied Physics} \bibinfo{volume}{102},
  \bibinfo{pages}{014504}.
%Type = Article
\bibitem[{Adachi(1988)}]{Adachi1988}
\bibinfo{author}{Adachi, S.}, \bibinfo{year}{1988}.
\newblock \bibinfo{title}{{Optical properties of AlxGa1-xAs alloys}}.
\newblock \bibinfo{journal}{Physical Review B} \bibinfo{volume}{38},
  \bibinfo{pages}{12345--12352}.
%Type = Book
\bibitem[{Adachi(1994)}]{Adachi1994}
\bibinfo{author}{Adachi, S.}, \bibinfo{year}{1994}.
\newblock \bibinfo{title}{{GaAs and Related Materials}}.
\newblock \bibinfo{publisher}{World Scientific}.
%Type = Article
\bibitem[{Losego et~al.(2009)Losego, Efremenko, Rhodes, Cerruti, Franzen and
  Maria}]{Losego2009}
\bibinfo{author}{Losego, M.D.}, \bibinfo{author}{Efremenko, A.Y.},
  \bibinfo{author}{Rhodes, C.L.}, \bibinfo{author}{Cerruti, M.G.},
  \bibinfo{author}{Franzen, S.}, \bibinfo{author}{Maria, J.P.},
  \bibinfo{year}{2009}.
\newblock \bibinfo{title}{{Conductive oxide thin films: Model systems for
  understanding and controlling surface plasmon resonance}}.
\newblock \bibinfo{journal}{Journal of Applied Physics} \bibinfo{volume}{106},
  \bibinfo{pages}{024903}.
%Type = Misc
\bibitem[{{Ioffe Physico-Technical
  Institute}()}]{IoffePhysico-TechnicalInstitute}
\bibinfo{author}{{Ioffe Physico-Technical Institute}}, .
\newblock \bibinfo{title}{{Electronic archive on semiconductor materials}}.
%Type = Article
\bibitem[{Sotoodeh et~al.(2000)Sotoodeh, Khalid and Rezazadeh}]{Sotoodeh2000}
\bibinfo{author}{Sotoodeh, M.}, \bibinfo{author}{Khalid, A.H.},
  \bibinfo{author}{Rezazadeh, A.A.}, \bibinfo{year}{2000}.
\newblock \bibinfo{title}{{Empirical low-field mobility model for III-V
  compounds applicable in device simulation codes}}.
\newblock \bibinfo{journal}{Journal of Applied Physics} \bibinfo{volume}{87},
  \bibinfo{pages}{2890--2900}.
%Type = Article
\bibitem[{Blandre et~al.(2017)Blandre, Chapuis and Vaillon}]{Blandre2017}
\bibinfo{author}{Blandre, E.}, \bibinfo{author}{Chapuis, P.O.},
  \bibinfo{author}{Vaillon, R.}, \bibinfo{year}{2017}.
\newblock \bibinfo{title}{{High-injection effects in near-field
  thermophotovoltaic devices}}.
\newblock \bibinfo{journal}{Scientific Reports} \bibinfo{volume}{7},
  \bibinfo{pages}{15860}.
%Type = Book
\bibitem[{Rytov et~al.(1989)Rytov, Kravstov and Tatarskii}]{Rytov1989}
\bibinfo{author}{Rytov, S.M.}, \bibinfo{author}{Kravstov, Y.A.},
  \bibinfo{author}{Tatarskii, Y.I.}, \bibinfo{year}{1989}.
\newblock \bibinfo{title}{{Principles of Statistical Radiophysics}}.
\newblock \bibinfo{publisher}{Springer-Verlag}.
%Type = Article
\bibitem[{Francoeur et~al.(2009)Francoeur, {Pinar Meng{\"{u}}{\c{c}}} and
  Vaillon}]{Francoeur2009}
\bibinfo{author}{Francoeur, M.}, \bibinfo{author}{{Pinar Meng{\"{u}}{\c{c}}},
  M.}, \bibinfo{author}{Vaillon, R.}, \bibinfo{year}{2009}.
\newblock \bibinfo{title}{{Solution of near-field thermal radiation in
  one-dimensional layered media using dyadic Green's functions and the
  scattering matrix method}}.
\newblock \bibinfo{journal}{Journal of Quantitative Spectroscopy and Radiative
  Transfer} \bibinfo{volume}{110}, \bibinfo{pages}{2002--2018}.
%Type = Article
\bibitem[{Callahan et~al.(2021)Callahan, Feng, Zhang, Toberer, Ferguson and
  Tervo}]{Callahan2021}
\bibinfo{author}{Callahan, W.A.}, \bibinfo{author}{Feng, D.},
  \bibinfo{author}{Zhang, Z.M.}, \bibinfo{author}{Toberer, E.S.},
  \bibinfo{author}{Ferguson, A.J.}, \bibinfo{author}{Tervo, E.J.},
  \bibinfo{year}{2021}.
\newblock \bibinfo{title}{{Coupled Charge and Radiation Transport Processes in
  Thermophotovoltaic and Thermoradiative Cells}}.
\newblock \bibinfo{journal}{Physical Review Applied} \bibinfo{volume}{15},
  \bibinfo{pages}{054035}.
%Type = Article
\bibitem[{Gummel(1964)}]{Gummel1964}
\bibinfo{author}{Gummel, H.K.}, \bibinfo{year}{1964}.
\newblock \bibinfo{title}{{A Self-Consistent Iterative Scheme for
  One-Dimensional Steady State Transistor Calculations}}.
\newblock \bibinfo{journal}{IEEE Transactions on Electron Devices}
  \bibinfo{volume}{11}, \bibinfo{pages}{455--465}.
%Type = Article
\bibitem[{Bernardi et~al.(2015)Bernardi, Dupr{\'{e}}, Blandre, Chapuis, Vaillon
  and Francoeur}]{Bernardi2015}
\bibinfo{author}{Bernardi, M.P.}, \bibinfo{author}{Dupr{\'{e}}, O.},
  \bibinfo{author}{Blandre, E.}, \bibinfo{author}{Chapuis, P.O.},
  \bibinfo{author}{Vaillon, R.}, \bibinfo{author}{Francoeur, M.},
  \bibinfo{year}{2015}.
\newblock \bibinfo{title}{{Impacts of propagating, frustrated and surface modes
  on radiative, electrical and thermal losses in nanoscale-gap
  thermophotovoltaic power generators}}.
\newblock \bibinfo{journal}{Scientific Reports} \bibinfo{volume}{5},
  \bibinfo{pages}{11626}.
%Type = Article
\bibitem[{Vaillon et~al.(2019)Vaillon, Perez, Lucchesi, Cakiroglu, Chapuis,
  Taliercio and Tourni{\'{e}}}]{Vaillon2019}
\bibinfo{author}{Vaillon, R.}, \bibinfo{author}{Perez, J.P.},
  \bibinfo{author}{Lucchesi, C.}, \bibinfo{author}{Cakiroglu, D.},
  \bibinfo{author}{Chapuis, P.O.}, \bibinfo{author}{Taliercio, T.},
  \bibinfo{author}{Tourni{\'{e}}, E.}, \bibinfo{year}{2019}.
\newblock \bibinfo{title}{{Micron-sized liquid nitrogen-cooled indium
  antimonide photovoltaic cell for near-field thermophotovoltaics}}.
\newblock \bibinfo{journal}{Optics Express} \bibinfo{volume}{27},
  \bibinfo{pages}{A11--A24}.
%Type = Article
\bibitem[{{El Oualid} et~al.(2021){El Oualid}, Kogut, Benyahia, Geczi, Kruck,
  Kosior, Masschelein, Candolfi, Dauscher, Koenig, Jacquot, Caillat, Alleno and
  Lenoir}]{ElOualid2021}
\bibinfo{author}{{El Oualid}, S.}, \bibinfo{author}{Kogut, I.},
  \bibinfo{author}{Benyahia, M.}, \bibinfo{author}{Geczi, E.},
  \bibinfo{author}{Kruck, U.}, \bibinfo{author}{Kosior, F.},
  \bibinfo{author}{Masschelein, P.}, \bibinfo{author}{Candolfi, C.},
  \bibinfo{author}{Dauscher, A.}, \bibinfo{author}{Koenig, J.D.},
  \bibinfo{author}{Jacquot, A.}, \bibinfo{author}{Caillat, T.},
  \bibinfo{author}{Alleno, E.}, \bibinfo{author}{Lenoir, B.},
  \bibinfo{year}{2021}.
\newblock \bibinfo{title}{{High Power Density Thermoelectric Generators with
  Skutterudites}}.
\newblock \bibinfo{journal}{Advanced Energy Materials} \bibinfo{volume}{11},
  \bibinfo{pages}{2100580}.
%Type = Phdthesis
\bibitem[{Blandre(2016)}]{Blandre2016}
\bibinfo{author}{Blandre, E.}, \bibinfo{year}{2016}.
\newblock \bibinfo{title}{{Thermal radiation at the nanoscale: near-field and
  interference effects in few-layer structures and on the electrical
  performances}}.
\newblock Ph.D. thesis. Universit{\'{e}} de Lyon.

\end{thebibliography}

\end{document}